\documentclass[ 10pt, journal,twocolumn,twoside]{IEEEtran} 
\usepackage{amsmath}
\allowdisplaybreaks[3]
\usepackage{graphicx}
\usepackage{epstopdf}
\usepackage{float}
\usepackage{array}
\usepackage{amsmath}
\usepackage{amssymb}
\usepackage{mdwmath}
\usepackage{CJK}
\usepackage{mdwtab}
\usepackage{eqparbox}
\usepackage{fixltx2e}
\usepackage{cases}
\usepackage{bm}
\usepackage{multirow}
\usepackage{psfrag}
\usepackage[usenames]{color}

\usepackage{amsmath}
\usepackage{subeqnarray}

\usepackage{fixmath}
\usepackage{mathdots}
\usepackage{latexsym}
\usepackage[numbers,sort&compress]{natbib}

\usepackage[ruled,linesnumbered]{algorithm2e}
\usepackage{url}
\ifCLASSOPTIONcompsoc
\usepackage[caption=false,font=normalsize,labelfont=sf,textfont=sf]{subfig}
\else
\usepackage[caption=false,font=footnotesize]{subfig}
\fi

\makeatletter
\renewcommand*{\@opargbegintheorem}[3]{\trivlist
	\item[\hskip \labelsep{\bfseries #1\ #2}] \textbf{(#3):}\ }
\makeatother

\begin{document}
	
	\title{Joint Service Caching, Communication and Computing Resource Allocation in Collaborative MEC Systems: A DRL-based Two-timescale Approach}

\author{\IEEEauthorblockN{
		Qianqian Liu,
		Haixia Zhang,~\IEEEmembership{Senior~Member,~IEEE,}  Xin Zhang and Dongfeng Yuan,~\IEEEmembership{Senior Member,~IEEE}}

}
	\maketitle
	\begin{abstract}
		Meeting the strict Quality of Service (QoS) requirements of terminals has imposed a significant challenge on Multi-access Edge Computing (MEC) systems, due to the limited multi-dimensional resources. 
%
		To address this challenge, we propose a collaborative MEC framework that facilitates resource sharing between the edge servers,  and with the aim to maximize the long-term QoS and reduce the cache switching costs through joint optimization of service caching, collaborative offloading, and computation and communication resource allocation. The dual timescale feature and temporal recurrence relationship between service caching and other resource allocation make solving the problem even more challenging.
		To solve it, we propose a deep reinforcement learning (DRL)-based dual timescale scheme, called DGL-DDPG, which is composed of a short-term genetic algorithm (GA) and a long short-term memory network-based deep deterministic policy gradient (LSTM-DDPG). 
		In doing so, we reformulate the optimization problem as a Markov decision process (MDP) where the small-timescale resource allocation decisions generated by an improved GA are taken as the states and inputted into a centralized LSTM-DDPG agent to generate the service caching decision for the large-timescale.
		Simulation results demonstrate that the proposed algorithm outperforms the baseline algorithms in terms of the average QoS and the cache switching costs.
		
	\end{abstract}
	\begin{IEEEkeywords}
		Multi-dimensional resources allocation, collaborative offloading, Service caching, two-timescale, long short-term memory (LSTM) network, deep deterministic policy gradient 
	\end{IEEEkeywords}
	
	\IEEEpeerreviewmaketitle

	\section{Introduction}
	
\IEEEPARstart{T}{he} evolution of the fifth generation mobile communication (5G), artificial intelligence (AI) and Internet of Things (IoT)  has facilitated the emergence of various intelligent applications in Industrial Internet of Things (IIoT) systems. Such are normally delay-sensitive and computation-intensive tasks, like AI-driven remote equipment monitoring, predictive maintenance, and quality control, which also have strict Quality of Service (QoS) requirements in terms of delay and energy consumption \cite{dai2023task, Chen2022}. Considering that the  time-varying sensing data and the low-power nature of  devices in IIoT environments, terminal devices (TDs) may not be able to support these demanding tasks alone. So, splitting and offloading tasks to the edge servers (ESs) is an alternative \cite{Qiu2020,Zhang2021b}.

With multi-access Edge Computing (MEC), the computing, storage, and intelligence resources can be sunk to the ESs, making the resources close to the TDs, thus can effectively reduce end-to-end delay and energy consumption needed for task processing \cite{Deng2022}.  A lot of works have  focused on how to schedule MEC's limited resources to meet the strict QoS requirements of the tasks. Most of works tackle this problem by optimizing the communication and computing resources to reduce  delay and energy consumption \cite{Liu2022a,Zhang2021d, Wang2022,Wu2022}. However, these studies have typically overlooked the fact that processing tasks offloaded from TDs may require specific service model stored at ESs ahead, including task-related machine learning models and databases known as edge service caching  \cite{Yan2021}. The edge service caching related to the short-term offloading and resource allocation, and the offloading strategy also depends on the service caching \cite{Chen2022a,Pham2021,Ko2022,Zhang2021f}.  Therefore, scheduling of multi-dimensional resources, i.e., the communication  computing and caching resources, should be done jointly at the ES, although they are in different timescales.


	Studies in \cite{Zhang2021a,Alqerm2021,Kamran2022} have investigated jointly allocating service caching, communication and computing resources in MEC networks. However, these works are done based on  an unrealistic assumption that service caching and other resource allocation decisions can be made simultaneously.  From the practical perspective, caching is a decision-making process based on statistical information over a large timescale, with decision-making intervals in days and hours. On the other hand, computing and communication involves instantaneous decisions made in seconds or minutes. If we optimize the multi-dimensional resources together on a small timescale, as in  \cite{Zhang2021a,Alqerm2021,Kamran2022}, will cause frequent cache switching and huge communication overheads. Therefore, for practical concern, the two should be jointly optimized over different timescales.

To address this, in this work  we consider a MEC-enabled IIoT system with the goal to improve the long-term QoS and reduce the cache switching costs. Achieving this  requires a reasonable resource allocation strategy for both large and small timescales. If we only rely on a given service caching strategy (large timescale), more than one ESs may cache the same service model. This will fail to utilize the storage diversity among ESs.  Besides, for the case that there is no service model at an ES, the tasks have to be offloaded to the  other ES or CS with the needed service model for  further processing. Such forward transmissions and results return will not only increase the communication overhead of the ES, but may also waste the computing resources of the ES, and finally  increase the task processing delay. 
On the contrary, if multiple ESs work together to cache the service models and handle the computation tasks cooperatively,  unnecessary transmission (tasks and service models) may be greatly reduced, and accordingly both TD's QoS and resource utilization can be improved  \cite{Kai2021a,Zhang2021,Sun2021,sahni2020multi,Tan2023,Chen2022b}.

	To the best knowledge of ours, few literature has addressed this in MEC-enabled IIoT systems. To fill this research gap, we investigate two-timescale multi-dimensional resources allocation with ESs collaboration to improve the long-term QoS and reduce the cache switching costs by jointly optimizing the long-term service caching, the short-term collaborative offloading, as well as the computing and bandwidth allocation.
	It is a typical  two-timescale optimization problem. How to handle the coupled multi-dimensional resources to realize the joint optimization of large and small timescales is the biggest challenge. Together with the collaboration among ESs make the problem even difficult to solve. 
To solve it, we utilize deep reinforcement learning (DRL) for the long-term service caching strategy, which is a powerful model-free learning method and is suitable for handling complex long-term sequential decision-making problems \cite{Wu2022,Zhang2021a,Alqerm2021,li2022mec}.  Additionally,  we use genetic algorithm (GA) to solve cooperative resource scheduling problems on small timescale, since it is easy to embed in DRL to enhance the training effect \cite{Zhang2022}.  More importantly, the long short-term memory (LSTM) network is used as a bridge between the
large and small timescales, leveraging its temporal processing and prediction capabilities to capture temporal relationships between the long-term service caching and the short-term offloading and resource allocation \cite{Zhang2023a, Li2022,Li2023,Zhang2021g,Liu2021}.
	Therefore, a two-timescale scheme based on GA, LSTM and DRL called DGL-DDPG is developed to address the joint cross-timescale multi-dimensional resources optimization problem. 
	For the large timescale, we propose a LSTM-assisted deep deterministic policy gradient (LSTM-DDPG) algorithm to generate service caching strategies. Meanwhile, for the small timescale, we propose an improved genetic algorithm (Improved-GA) to generate immediate decisions for collaborative offloading, computing, and bandwidth resource allocation.
	The main contributions of the paper can be summarized as follows:	
	\begin{enumerate}
		\item 
		We establish a collaborative MEC-enabled IIoT system architecture in which the ESs  collaborate with each other to cache service models, and handle computational tasks.  We propose a two-timescale multi-dimensional resources scheduling scheme, named as DGL-DDPG, to jointly optimize the long-term service caching, short-term collaborative offloading, computing, and bandwidth resource allocation  for the established system to maximize the long-term QoS of all TDs and reduce the long-term cache switching costs of all ESs.
		\item DGL-DDPG can intelligently and  dynamically schedule multi-dimensional resources across timescales with the assistance of  the LSTM-DDPG and the Improved-GA.  In doing so, we transform multi-dimensional resource allocations at different timescales into a Markov Decision Process (MDP), where instantaneous resource allocation decisions generated by  Improved-GA are treated as states. By feeding these states into a centralized LSTM-DDPG agent, the long-term service caching decisions are generated.
		\item The LSTM-DDPG based service caching algorithm can model the temporal relationship between the long-term service caching and the short-term offloading and predict service demand variance, and make  a long-term optimal service caching decision.
		Furthermore, we  design  a hybrid coding based Improved-GA to generate both discrete offloading decisions and continuous allocation of computing and bandwidth resources at small-timescale.	
	\end{enumerate}

	The remainder of this paper is organized as follows. Section II presents the related work. Section III describes the system model and problem formulation. Section IV elaborates on the DRL-based two timescale optimization algorithm. Section V presents simulation results that evaluate the convergence and performance of the proposed algorithms. Finally, Section VI presents the conclusions drawn from the study.

	\section{Related Work}
	This section provides a review of recent studies on resource allocation in MEC systems, categorized based on joint allocation of communication and computing  resources, joint allocation of service caching and  computing resources, and joint allocation of service caching, communication and computing  resources.
	
	\subsection{Joint communication and computing resources allocation}  
	For the allocation of communication and computing resources in MEC networks, most researcher focus on reducing the task processing delay and/or energy consumption through aspects such as task offloading (partially), computing resource allocation, power control, and channel selection.
	
	\citet{Tan2022} proposed a two-level alternation method to minimize total energy consumption in multi-user collaborative MEC networks. The method uses a heuristic algorithm in the upper level for collaboration and offloading decisions and a DRL-based algorithm in the lower level for power, subcarrier, and computing resource allocation.
	\citet{Kai2021} proposed a pipeline-based offloading scheme  to minimize the overall latency for all users in a could-edge-end collaborative MEC network. They utilized the classic successive convex approximation (SCA) approach to optimize the offloading strategy, computation resource allocation, delivery rates, and power allocation. 
	In \cite{Deng2022a}, the authors considered the increasing tasks with stringent QoS requirements, as well as the limited communication and computational resources in multi-user multi-edge MEC networks. They formulated a long-term throughput maximization problem from the perspective of service providers under latency constraints.
	However, the above literature does not take into account the impact of task-related service caching. They all take it for granted that all the needed services are  cached at all ESs. This is not realistic for the ESs with limited cache resources.
	
	\subsection{Joint service caching and computing resources allocation}
	Works on service caching configuration and computing offloading mainly aim to reduce the task processing delay and energy consumption by jointly scheduling  the services models, offloaded tasks, and computing resources within a small-timescale slot.
	
	With the aim of maximizing the users' QoS, \citet{Chu2022} proposed a Lagrangian-dual based distributed algorithm to optimize the service selection and computing resource allocation for a single MEC network. They further designed an online service caching update strategy based on the cost of services switching. Similarly, \citet{Zhou2022a} extended Chu’s optimization problem to a multi-ES collaborative MEC system and proposed a DDPG-based algorithm to minimize the long-term energy consumption.
	\citet{Zhou2023} investigated a use case of reverse computation offloading and service caching in a three-tier MEC system, in
	which the  cloud  server  can purchase computation and communication resources from the ESs to assist with computing the offloaded tasks.
	Generally, the caching resources are adjusted at a larger timescale than that the computation resources are done at. However, this feature has not been captured in the literature mentioned above.
	For this, \citet{Zhang2023} proposed a hierarchical DDPG-based algorithm to deal with the dual timescale resource allocation for IoT systems to minimize the delay and energy consumption. 
	Lyapunov optimization approach and dependent rounding technique are adopted  to solve the problem of the long-term service caching and the short-term UAV trajectory, user association, and offloading decisions optimization, respectively \cite{Zhou2022b}.
	However, the aforementioned literature has not taken into account  the communication resource needed in the caching and offloading process.
	
	\subsection{Joint service caching, communication and computing resources allocation} 
	For the joint multi-dimensional resources optimization in MEC networks, most literature mainly studies service caching configuration, task offloading, and computing and communication resources allocation at the same timescale, with the goal of reducing the task processing delay, energy consumption, and resource purchase costs.
	
	\citet{Bi2020} have investigated a single MEC scenario where a user has a series of dependent tasks. They transformed the joint optimization problem of the service caching placement, computation offloading, and system resource allocation ( CPU processing frequency and transmit power of user) into an equivalent pure 0-1 integer linear programming for solution.
	In order to minimize the computation and delay costs, \citet{Zhang2021a} investigated the joint allocation of service caching, transmission, and computing resources in a general scenario of multiple users with multiple tasks. They also considered the scenario where each user has a computation cost budget and formulated  a quadratically constrained quadratic program problem solved with a semi-definite relaxation-based algorithm. 
	However, these studies  make the ideal assumption that caching and other resources can be scheduled simultaneously, which leads to frequent cache switching and high system communication overheads.
	
	Several researchers, including \citet{Wen2020} and \citet{Sun2022}, have addressed the issue by incorporating dual timescale into account in doing multi-dimensional resources allocation. In \cite{Wen2020}, a consensus alternating direction method of multipliers-based joint policy for the caching, offloading, and time slot allocation was proposed to minimize the weighted sum energy consumption subject to the storage and delay constraints. However, the temporal relationship between the caching and the offloading is replaced by simple statistics, meaning that the caching strategy is done based solely  on the static statistics. To address this, \citet{Sun2022} proposed a joint multi-agent RL and game approach to manage multi-dimensional resources in Fog radio access networks. RL is used to learn the long-term service caching strategies from historical information of user requests and channel states in  small timescales. However, due to the inherent limitations of RL, the temporal relationship between the caching and the offloading is still not fully utilized.
	In addition, these works generally focus on the resource allocation of a single ES, while ignoring the collaboration of multiple ESs. 
	Therefore, how to design a intelligent   two-timescale  multi-dimensional resources allocation scheme with collaborative ESs remains a challenging problem.
	To address this problem, we establish a collaborative MEC-enabled IIoT system, and propose a DRL-based two-timescale multi-dimensional resources scheduling  scheme, DGL-DDPG.

	\begin{figure}[t]
		\centering
		\includegraphics[width=0.5\textwidth]{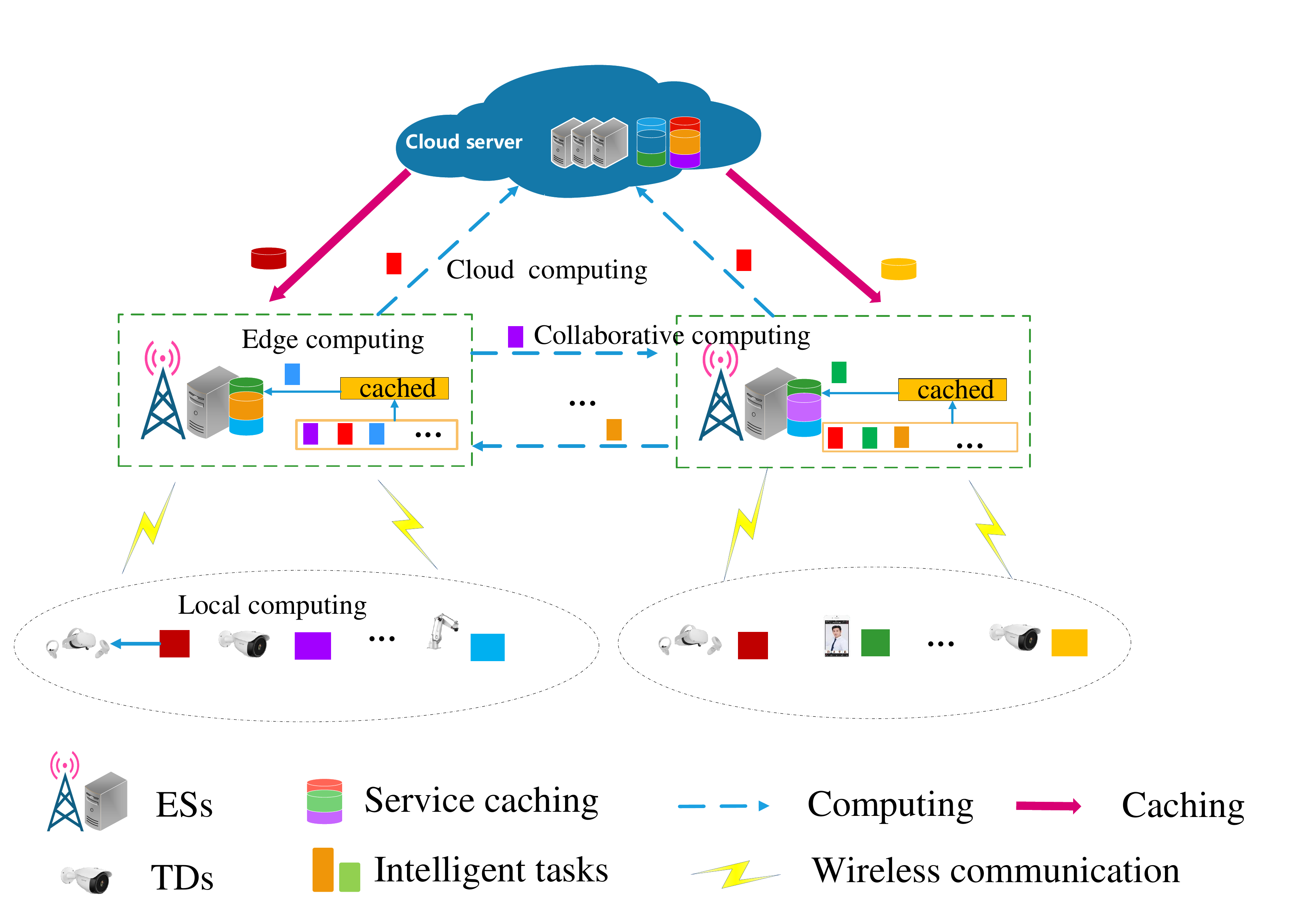}\\
		\caption{Illustration of service caching, communication and computing in a collaborative MEC-enabled IIoT system.}
		\label{systemmodel}
	\end{figure}
	
	\section{System Model and Problem Formulation }

	\subsection{Network Architecture}
	As shown in Fig. \ref{systemmodel}, the collaborative MEC network comprises of one cloud server (CS), $M$ ESs, and $N$ TDs. 
	Each TD can connect with one and only one ES via a wireless link at a time, and the ES can communicate with other ES and CS through wired links. The ES $m$ can act as an access point (AP) with a built-in MEC server, which has limited available spectrum, computing, and caching resources, represented by $\{W_m,C_m,F_m\}$. We assume that all ESs can be regarded as collaborative servers and are willing to share their computing and caching resources with each other.
	There are $F$ types of computing tasks and corresponding $F$ types of service models. The service set is denoted as $\mathcal{F}=\{1,2,\cdots,f,\dotsb,F\} $. 
	We adopts a discrete time model which is divided into two-timescale,
	i.e., large-timescale for service caching placement and small-timescale
	for task ofﬂoading and resource allocation as shown in Fig.\ref{two-timescale}. The large-timescale index of caching is denoted by
	$i \in I = \{1, 2, ..., I\}$, each of which includes $T$ short timescale
	slots. The small-timescale index  within the $i$th large-timescale slot is represented by $k \in K = \{1, 2, ..., K\}$. At each short time slot $k$, TD $n$ randomly generates a computation task. We use a three-tuple 
	$\{s_{n,i,k}^{type},s_{n,i,k}^{input},s_{n,i,k}^{comp}\}$  to denote the task of the $n$th TD, where $s_{n,i,k}^{type}\in \mathcal{F}$ represents the type of the task,  $s_{n,i,k}^{input}$ represents the amount of input data (in bits), and $s_{n,i,k}^{comp}$ represents the required computational capacity (in CPU cycles). 
	AS illustrated in Fig. \ref{systemmodel}, the tasks of TDs can be processed locally (e.g., red task) or offloaded to the ESs for computing. When the task is offloaded to the associated ES, if the corresponding service is cached, the task will be executed at the ES (e.g., blue and green tasks). Alternatively, the task can be further offloaded to the other collaborative ES for computing (e.g., purple and orange tasks). If none of the collaborative ES have cached the service, the task will be executed at the CS (e.g., red tasks). At each  large-timescale slot $i$, the ESs update service caching placement (e.g., red and orange service models).

	\subsection{Service Caching Model}
	We define $\mathbf{c}(i) = \{c_{m}^f(i)\in \{0,1\}| m\in \mathcal{M},f\in\mathcal{F}\}$ to denote the service caching decision of all ES at large-timescale $i$ and  the cache conﬁgurations of all ES maintain unchanged for a large-timescale slot. If the ES $m$ cached the $f$th service at the $i$th large-timescale, $c_{m}^f(i) =1$, otherwise, $c_{m}^f(i) = 0$. 
	The total cached services should not exceed the cache capacity
	of the ESs, that is
	\begin{equation}\label{eq1}
		\sum_{f\in\mathcal{F}}{c_{m}^f(i) s_{f}^{cache}}\le C_m, \forall m\in\mathcal{M}, \forall i\in\mathcal{I},
	\end{equation}
	where $S^{cache}_f$ is the storage size of the $f$th service model (in
	bits). Services in $F$ are all available at the CS and can be
	cached to the ESs via backhaul. The cache switching cost is defined as the communication delay cost for downloading the services,
	\begin{equation}\label{eq2}
		Cost(i) =  \sum_{m=1}^M\sum_{f=1}^F\mathbb{I}[(c_f^i-c_f^{i-1}) =1] \frac{s_f^{cache}}{R^{back}},
	\end{equation} 
	where 
	$\mathbb{I}[x] $ is an indicator function that equals 1 if condition $x$ is true and 0 otherwise, $(c_m^f(i)-c_m^f(i-1)) =1$ indicates that the service should be obtained from the CS, $ R^{back}$ is the fixed  backhaul transmission rate. Obviously, the cost is 0 at  the small-timescale slots.

	\subsection{Computing Model}
	We define $\mathbf{d}(k) = \{d_{n,i}^m(k)\in\{0,1\}| n\in \mathcal{N},m\in \mathcal{M}\cup\{0,C\}\}$ to be the decision variable for TDs' collaborative offloading at the $k$th small-timescale slot in the $i$th large-timescale slot, $d_{n,i}^m(k) = 1$ indicates that the $n$th TD  offloads its task to  the $m$th ES at time slot $k$, similarly, $d_{n,i}^0 (k) = 1$ indicates  local computing and $ d_{n,i}^C(k) = 1$ indicates cloud computing. Besides, we define $\mathbf{f}(k)=\{f_{n,i}^m(k) \in [0,1]|n\in \mathcal{N},m\in \mathcal{M} \}$ as the  computing resource allocation decision,  $f_{n,i}^m (k) F_m$ represents the computing frequency allocated to TD $n$ by ES $m$.  It should be noted that the task can only be processed at the
	ES if and only if the relevant service model is cached at the ES
	and the necessary computing resources are allocated to it too. Hence,
	we have 
	\begin{equation}\label{eq2}
		d_{n,i}^m (k) \le \lceil f_{n,i}^m (k) \rceil \le c_{m}^f (i), \forall n \in \mathcal{N},\forall m \in \mathcal{M}.  
	\end{equation}
	
	According to different collaborative offloading decisions, the computing mode of TD $n$ can be divided into four cases:
	
	Case 1: Local computing only ie., $d_{n,i}^0(k) = 1$. The local computing delay (in seconds ) is
	\begin{equation}\label{eq3}
		T_{n,i}^{local}(k) = \frac{s_{n,i,k}^{comp}}{F_n^{l}},
	\end{equation}
	where $F_n^l$ is the computing capability of TD $n$. The corresponding energy consumption is 
	\begin{equation}\label{eq4}
		E_{n,i}^{local}(k) =  \varepsilon (F^l_n)^2 s_{n,i,k}^{comp},
	\end{equation}
	where $\varepsilon$ is the energy coefficient depending on the chip architecture of the TD $n$. 
	
	Case 2: Edge computing at associated ES $m$.  In this case, the task is offloaded to ES $m$ and the service model of task is cached, ie., $d_{n,i}^m(k) = 1$ and $ c_{m}^{f}(i) = 1$. Then the computing latency can be given as
	\begin{equation}\label{eq5}
		T_{n,m,i}^{comp} (k)= \frac{s_{n,i,k}^{comp}}{f_{n,i,k}^{m}F_m}.
	\end{equation}
	
	Case 3: Edge computing at cooperative ES $\acute{m}$.  In this case, there is no ability to process the offloaded task at associated ES and the task is forwarded to the cooperative ES $\acute{m}$ for processing, where the service model of task is cached  and  the resource is sufficient  at ES $\acute{m}$, ie., $d_{n,i}^{\acute{m}} (k)= 1\  \mathbf{and}\  c_{\acute{m}}^{f}(i) = 1$. Then the computing latency can be express as
	\begin{equation}\label{eq6}
		T_{n,\acute{m},i}^{comp}(k) = \frac{s_{n,i,k}^{comp}}{f_{n,i}^{\acute{m}}(k)F_{\acute{m}}},
	\end{equation}
	where $ \acute{m} \in \mathcal{M} / m $ is the cooperative ES of the ES $m$, $f_{n,i}^{\acute{m}}(k)F_{\acute{m}}$ is the computing frequency allocated to TD $n$ by ES $\acute{m}$.
	
	Case 4: Could computing. In this case, the service model of task is not cached at any ES, so it can only be further offloaded to the CS for computing,  ie., $d_{n,i}^C(k) = 1$. The computing latency can be written as 
	\begin{equation}\label{eq7}
		T_{n,C,i}^{comp}(k)  = \frac{s_{n,i,k}^{comp}}{F_{c}},
	\end{equation}
	where $F_c$ is the computing capability of the CS.
	
	\subsection{Communication Model}
	Considering that the data size of the task processing results is negligible compared to the data size of the tasks themselves, in this work we only focus on the uplink transmission. We consider the scenario that each ES is pre-assigned spectrum bandwidth to serve its associated TDs, and we define $\mathbf{b}(k) = \{b_{n,i}^m(k) \in [0,1]|n\in \mathcal{N},m\in \mathcal{M}\}$ as the bandwidth allocation decision, $b_{n,i}^m (k) $ represents the bandwidth allocated to the $n$th TD by  the $m$th ES for task offloading at the $k$th small-timescale slot within the $i$th large-timescale slot. 
	The uplink transmission rate from TD $n$ to ES $m$ can be written as 
	\begin{equation}\label{eq8}
		R_{n,i}^{m}(k) = b_{n,i}^m(k) W_m log_2((1+\frac{p_n h_{n,m,i}(k)}{\sigma^2}),
	\end{equation}
	where $p_n$ is the uplink transmit power of TD $n$ , $h_{n,m,i}(k) = g_{n,m,i}(k)dis_{n,m}^{-\alpha} $ represents the corresponding channel gain, with $g_{n,m}(k)$ and $dis_{n,m}^{-\alpha} $ are the small-scale Rayleigh fading and the large-timescale path fading, respectively, $dis_{n,m}$ represents the distance between TD $n$ and  ES $m$ and $\alpha$ is the path loss exponent of the TD-ES links, $\sigma^2 $ denotes the variance of the additive white Gaussian noise. We assume that the wired links transmit data at a fixed rate $R^{co}$ and $R^{back}$ (in bits per second, bps).
	The transmission delay and corresponding energy consumption from TD $n$ to its associated ES $m$ for task offloading can be defined as
	\begin{equation}\label{eq9}
		T_{n,m,i}^{trans}(k) = \frac{s_{n,i,k}^{input}}{R_{n,i}^{m}(k)}.
	\end{equation}
	\begin{equation}\label{eq10}
		E_{n,i}^{trans}(k) =p_n \times T_{n,m,i}^{trans}(k).
	\end{equation}
	The transmission delay from ES $m$ to the cooperative ES $\acute{m}$ for task offloading is written as 
	\begin{equation}\label{eq11}
		T_{m,k,i}^{trans}(k) = \frac{s_{n,i,k}^{input}}{R^{co}}.
	\end{equation}
	The transmission delay from ES $m$ to the CS for task offloading is written as 
	\begin{equation}\label{eq12}
		T_{m,C,i}^{trans}(k) = \frac{s_{n,i,k}^{input}}{R^{back}}.
	\end{equation}
	\subsection{QoS model}
	The delay for processing the task of the $n$th TD at the $k$th small-timescale slot within the $i$th large-timescale slot for any case
	can be expressed as
	\begin{small}
	\begin{equation}\label{eq13}
		T_{n,i}(k) = 
		\begin{cases}
			T_{n,i}^{local}(k), & \text{$d_{n,i}^0(k) = 1$}\\
			T_{n,m,i}^{trans}(k) + T_{n,m,i}^{comp}(k),& \text{$d_{n,i}^m(k) = 1$}\\
			T_{n,m,i}^{trans}(k) +T_{m,\acute{m},i}^{trans}(k)+ T_{n,\acute{m},i}^{comp}(k),&\text{$d_{n,i}^{\acute{m}}(k) =1$}\\
			T_{n,m,i}^{trans}(k) +T_{m,C,i}^{trans}(k)+ T_{n,C,i}^{comp}(k),&\text{$d_{n,i}^C(k) =1$}
		\end{cases}.
	\end{equation}
	\end{small}
	The energy consumption for processing the task of the $n$th TD at the $k$th small-timescale slot within the $i$th large-timescale slot for any case can be expressed as
	\begin{equation}\label{eq14}
		E_{n,i}(k) = (1-d_{n,i}^0(k)) E_{n,m,i}^{trans}(k) + d_{n,i}^0(k) E_{n,i}^{local}(k).
	\end{equation}
	
	As stated in Section I, delay and energy consumption are critical performance metrics for TDs in IIoT systems. Therefore, we define the QoS for TDs, in terms of the satisfaction with latency and energy consumption relative to a set threshold. This can be expressed as
	\begin{equation}\label{eq16}
		Q_{n,i}(k) = \alpha \times\frac{T_{n}^{th}- T_{n,i}(k)}{T_{n}^{th}}  + 
		\beta \times\frac{E_{n}^{th}- E_{n,i}(k)}{E_{n}^{th}} ,
	\end{equation}
	where $\alpha,\beta \in[0,1]$, $\alpha +\beta = 1$ are the weighted parameters specifying the TD's preferences for delay and energy consumption in the QoS, respectively, $T_{n}^{th},E_{n}^{th}$  are the threshold of the  latency and energy, which guarantee TD $n$ can work.
	
	\subsection{Problem Formulation }\label{Problem Formulation}
	Our objective is to maximize the long-term QoS of all TDs, while reducing the long-term cache switching costs of all ESs  in a collaborative MEC-enabled IIoT system. To achieve this goal, we propose a two-timescale multi-dimensional resource joint optimization framework by properly designing the service caching, collaborative offloading, computation, and bandwidth resource allocation. The problem of multi-dimensional resource allocation can be formulated as
	\begin{alignat}{2}
		\max_{\mathbf{c}(i),\mathbf{d}(k),\mathbf{f}(k),\mathbf{b}(k)}\quad &\lim_{I\rightarrow\infty}\frac{1}{I}\sum_{i=1}^{I}[\sum_{k=1}^{K}\sum_{n = 1}^{N} 
		Q_{n,i}(k) - \delta Cost(i)]\\
		\mbox{s.t.} \quad 
		& d_{n,i}^m(K) \le \lceil f_{n,i}^m(K) \rceil \le c_{m}^f(k),\nonumber \\ 
		&\ \ \ \ \ \ \ \ \ \ \ \ \ \ \ \ \ \forall n \in \mathcal{N},\forall m \in \mathcal{M} &\tag{17a}\\
		&b_{n,i}^m(k)\le (1-d_{n,i}^0(k)),  \forall n\in\mathcal{N},\nonumber \\
		&\ \ \ \ \ \ \ \ \ \ \ \ \ \ \ \ \ \ \ \ \ \ \ \ \ \  \ \forall m\in\mathcal{M}&\tag{17b}\\
		&\sum_{m\in \mathcal{M}\cup\{0,C\}}d_{n,i}^m(k) = 1,\forall n\in\mathcal{N} &\tag{17c}\\
		&\sum_{f\in\mathcal{F}}{c_{m}^f(i)s_{f}^{cache}}\le C_m, \forall m\in\mathcal{M}&\tag{17d}\\
		&\sum_{n\in\mathcal{N}}{d_{n,i}^m(k) f_{n,i}^m(k)}\le 1, \forall m\in\mathcal{M}&\tag{17e}\\
		&\sum_{n\in\mathcal{N}}{b_{n,i}^m(k)}\le 1, \forall m\in\mathcal{M}&\tag{17f}\\
		&E_{n,i}(k)\le E_{n}^{th}, \forall n\in\mathcal{N}&\tag{17g}\\
		&T_{n,i}(k)\le T_{n}^{th}, \forall n\in\mathcal{N}&\tag{17h}\\
		& c_{m}^f(i) \in\{0,1\}, \forall m\in\mathcal{M}, \forall f\in \mathcal{F}&\tag{17i}\\
		&d_{n,i}^m(k) \in \{0,1\},\forall n\in\mathcal{N}, \nonumber \\
		&\ \ \ \ \ \ \  \ \ \ \ \ \ \ \ \ \ \ \ \forall m\in\mathcal{M}\cup\{0,C\}&\tag{17j}  \\
		&f_{n,i}^m(k)\in[0,1], \forall n\in\mathcal{N}, \forall m\in\mathcal{M}&\tag{17k}\\
		&b_{n,i}^m(k) \in[0,1],\forall n\in\mathcal{N},\forall m\in\mathcal{M}&\tag{17m}
	\end{alignat}
	where $\delta$ is a factor balancing the weight of the  cache switching cost and the TDs'QoS,  the constraint
	(17a) indicates that the task cannot be processed if there is
	no relevant service model cached at the ES or no computing
	resources have been allocated to it,
	Constraint (17b) implies that the bandwidth of ESs is allocated only to the TDs offloading their tasks, Constraint (17c) indicates that the task generated by TD $n$ can only be processed by a single server, Constraints (17d, 17e, 17f) denote the limitations of the caching, computing, and bandwidth capacities of each ES, respectively, Constraints (17g, 17h) indicate the maximum delay and energy limitations of TDs. 
	Constraints (17i, 17j, 17k, 17m) make the problem a mixed-integer nonlinear programming (MINLP) problem, with variables on different timescales, which typically results in an NP-hard problem. The challenge lies in the difficulty of modeling the temporal relationship between the service caching and the task offloading. The scale and complexity of the problem are prohibitively high for utilizing traditional optimization methods to solve it. To address this issue, we propose a two-timescale solution DGL-DDPG in the next section.

	\section{ DRL-based Two-timescale Resource allocation}\label{ Algorithm Design}
	In this section, we first introduce the overall framework of the designed two-timescale multi-dimensional algorithm DGL-DDPG, and then we  introduce the large-timescale LSTM-DDPG-based service caching algorithm  and the Improved-GA-based collaborative offloading, computing and bandwidth resource allocation  algorithm  for the small timescale. Finally, the complexity  of the proposed DGL-DDPG  is analyzed. 
	
	\begin{figure*}[t]
		\centering
		\includegraphics[width=0.95\textwidth]{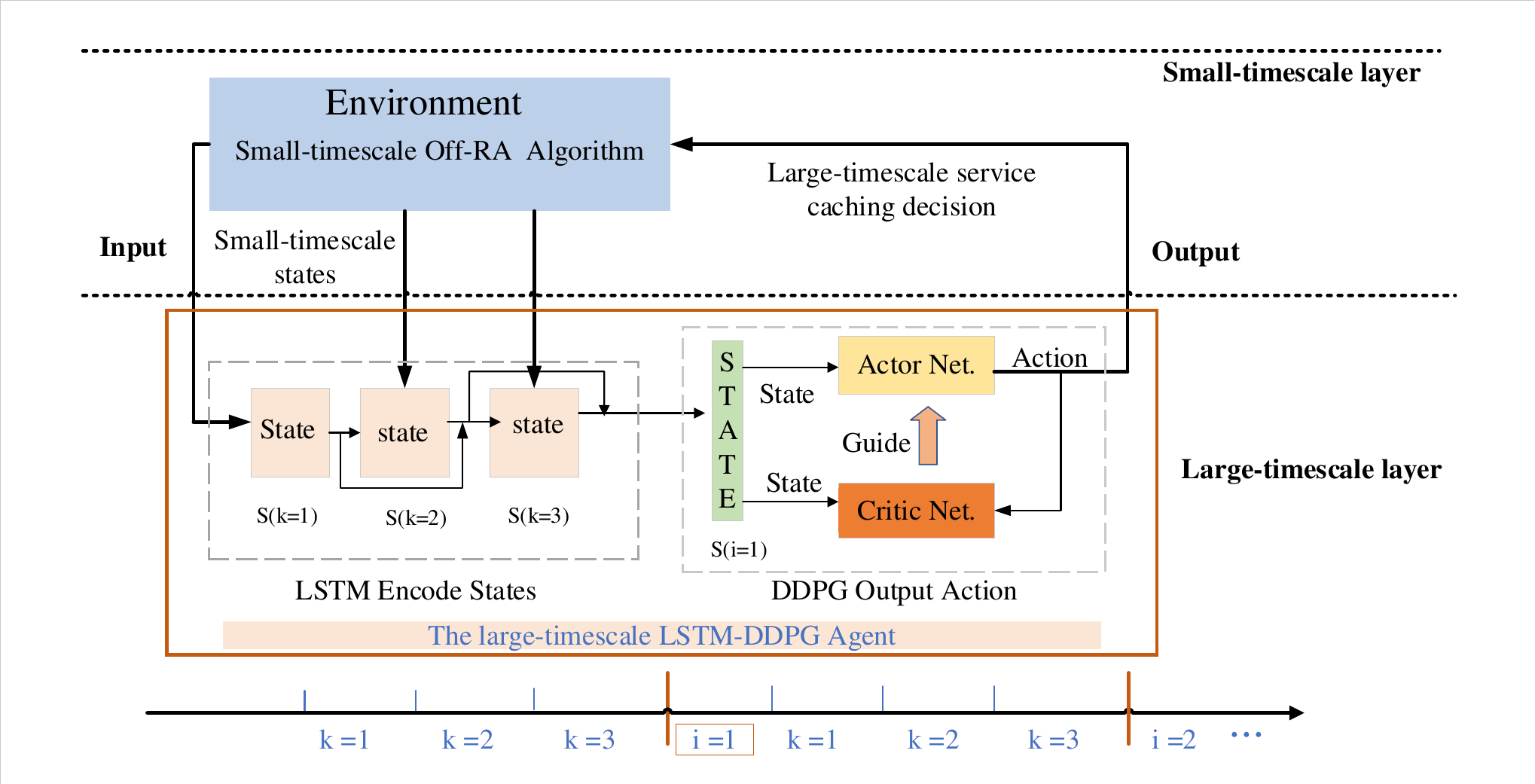}\\
	\caption{The framework of the DGL-DDPG. }\label{two-timescale}
		
	\end{figure*}

	\begin{figure*}[t]
		\centering
		\includegraphics[width=0.95\textwidth]{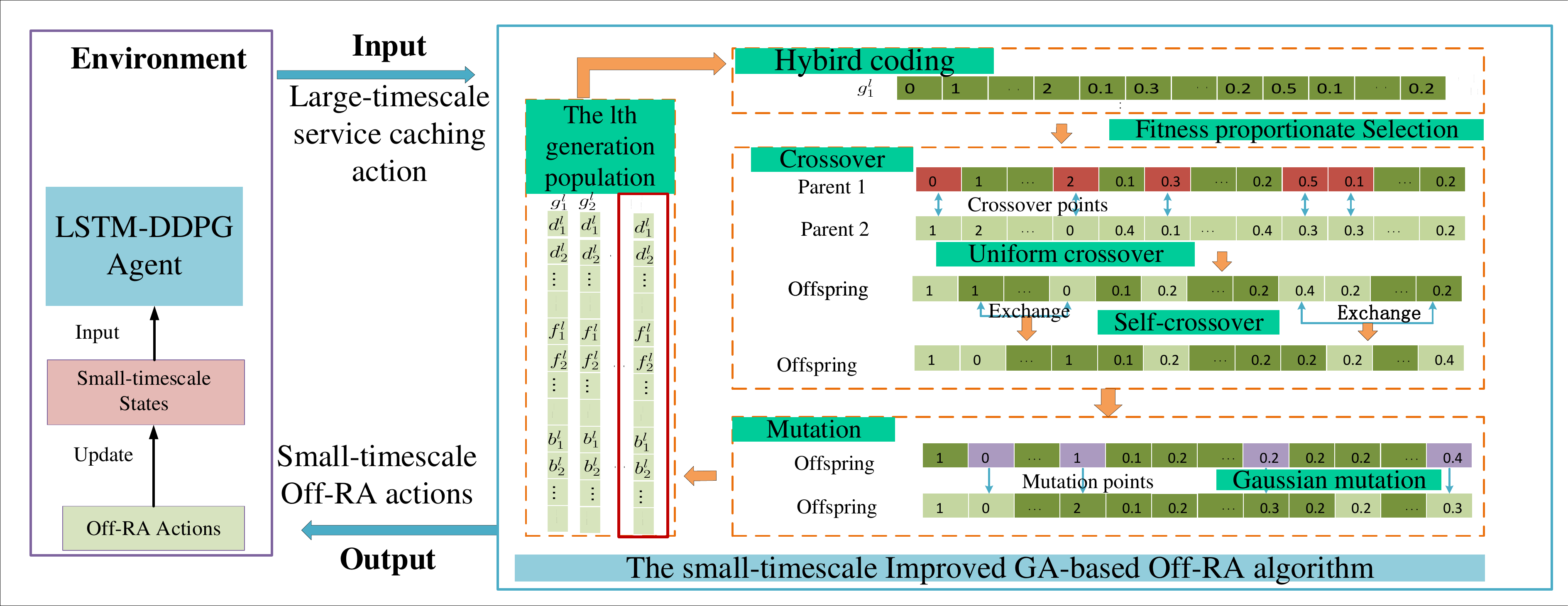}\\
		\caption{The framework of the small-timescale Improved GA based Off-RA. }\label{two-timescale2}
		
	\end{figure*}

	\subsection{The framework of the DGL-DDPG-based two-timescale algorithm }\label{cache}
As illustrated in Fig. \ref{two-timescale},  the DGL-DDPG algorithm  consists of two layers, one is the large-timescale LSTM-DDPG based service caching, the other is the small-timescale Improved-GA based  Off-RA. Specifically, service caching decisions are made at the beginning of each large-timescale slot $i$ by an LSTM-DDPG agent, while each large-timescale caching slot is composed of $K$ small-timescale Off-RA slots. The Improved-GA  is used to make the corresponding  Off-RA decisions in each small-timescale slot $k$.  
In particular, Fig. \ref{two-timescale} shows the case that  a large-timescale  slot includes three small-timescale slots. First of all, the LSTM-DDPG agent  observes the states of the environment over the first three small-timescale slots and feeds them into the LSTM network to learn,
and then the LSTM  network outputs an encoded state, that is, the current state of the large-timescale slot. It is then fed into DDPG network to get  the large-timescale service caching decision. Finally, the caching decision is executed in the environment, the small-timescale Improved-GA will continue to output the Off-RA decisions and update the states of the environment  under the new caching decision, and the  LSTM-DDPG agent will continue to collect states and update the caching decision at the next large-timescale slot. The core of the two-layer optimization is to use the LSTM network to predict the time-varying cache requirements on a small timescale and, similar to the iterative algorithm, the two layers of DGL-DDPG are trained together until convergence. Correspondingly, as shown in Fig.  \ref{two-timescale2}, the input of the small-timescale  Improved-GA algorithm is the service caching decision, and the output is the Off-RA decisions.

	\subsection{LSTM-DDPG based Service Caching Strategy   }\label{cache}
The variant Off-RA decisions affect the large-timescale service caching decisions, and the current decisions will also affect the future decisions. For example, the ESs tend to store those services that are frequently used but not cached. So, this is a typical dynamic sequential decision problem. Based on this, the dynamic environment and original optimization problem can be transformed into a Markov decision process (MDP) across different timescales and then we use deep reinforcement learning algorithm to solve it.	
	The reason for choosing LSTM-DDPG as the DRL agent mainly because of its capabilities of model-free temporal decision making and state prediction. Specifically, DDPG is a state-of-the-art DRL algorithm based on stable and efficient actor-critic framework. When the LSTM network is embedded in DDPG as the temporal feature extraction network future caching demands can be predicted. In the following, we define the state, action and reward function of service caching and then the design of LSTM-DDPG agent was presented.
	
	\emph{1) Service caching MDP}.
	\textbf{State}  Since the service caching strategy relies heavily on the small-timescale Off-RA decisions and collaboration between ESs. The state space consists of the  current service caching conﬁguration of each ESs, the cumulative cache gain of each service, and  the popularity of each service on each ES and is written as  $\mathbf{s}_{k} = \{\mathbf{c}_{k},\mathbf{g}_{k},\mathbf{p}_{k}\}$, with
	
	

	\begin{equation}\label{eq17}
		\mathbf{c}_{k} = \{c_{m}^f(k)| m\in \mathcal{M},f\in\mathcal{F}\}_{M\times F},
	\end{equation} 
	
	\begin{equation}\label{eq18}
		\mathbf{g}_{k} = \{g_{f}(k) | f\in\mathcal{F} \}_{1\times F},
	\end{equation}
where $g_{f}(k)$  describes the cumulative cache gain of service $f$, $g_{f}(k) = \sum_{\tau \in I} \sum_{k\in K} \sum_{n\in N} \mathbb{I}[s_{n,\tau,k}^{type} =f] Q_{n,\tau}(k)-Q_{n,\tau}^l(k)$, and $Q_{n,\tau}(k)-Q_{n,\tau}^l(k)$ represents the QoS gain caused by edge service  caching compared to local computing, $Q_{n,\tau}^l(k) = \alpha \times\frac{T_{n,\tau}^{th}(k)- T_{n,\tau}^{l}(k)}{T_{n,\tau}^{th}(k)}  + \beta \times\frac{E_{n,\tau}^{th}(k)- E_{n,\tau}^l(k)}{E_{n,\tau}^{th}(k)}$  denotes the Qos of TD $n$ when computing locally.

	\begin{equation}\label{eq19}
		\mathbf{p}_{k}=\{p_{m,f}(k)| m\in \mathcal{M},f\in\mathcal{F}\}_{M\times F}. 
	\end{equation} 
where $p_{m,f}(k) = \sum_{\tau \in I} \sum_{k\in K} \sum_{n\in N} \mathbb{I}[s_{n,\tau,k}^{type} =f]d_{n,\tau}^m(k)$  represents the popularity of service $f$ in ES $m$.
		

	To predict the time-varying service caching demands , the agent will observe the state once in each small-timescale slot $k$, and then use the LSTM network to transform all the small time slot states$\{s_1,\cdots,s_k,\cdots,s_T\}$ into a large-timescale state $S_i$, which will be feed into DDPG network to output the large-timescale service caching decision.
	\begin{equation}\label{eqlstm}
		S_i = LSTM(s_1,\cdots,s_k,\cdots,s_T)
	\end{equation}

	\textbf{Action:} The action is the service caching policy of each ES, $\mathbf{a}_{i} = \{\mathbf{c}_{1}(i),\cdots,\mathbf{c}_{M}(i)\}_{M\times F},\mathbf{c}_{m}(i) = \{c_{m}^1(i),\cdots,c_{m}^f(i),\cdots,c_{m}^{F}(i)\}$.
	
	\textbf{Reward:} For better caching strategy, we take into account the  long-term QoS of TDs,  the cache switching cost and  the penalty for exceeding the cache space. The reward function can be described as follows.
	\begin{equation}\label{eq11}
		r_{i} = r_{qos}(i) - r_{cost}(i)- r_{penalty}(i).
	\end{equation} 
	\begin{equation}\label{eq11}
		r_{qos}(i) = \sum_{k= 1}^{K}\sum_{n=1}^{N} Q_{n,i}(k).  
	\end{equation} 
	\begin{equation}\label{eq11}
		r_{cost}(i) = \sum_{m=1}^{M} \sum_{f=1}^{F}\mathbb{I}[c_{m}^f(k)-c_{m}^f(k-1) =1] \frac{s_f^{cache}}{R^{back}}.
	\end{equation} 
	\begin{equation}\label{eq11}
		r_{penalty}(i) =  \sum_{m=1}^{M} \min\{(\sum_{f=1}^{F} c_{m}^f(k) s_f^{cache} - C_m),0\}. 
	\end{equation} 
	\emph{2) The design of LSTM-DDPG}.
	As illustrated in Fig. \ref{two-timescale}, LSTM-DDPG comprises three main components: the actor network, the critic network, and an episode experience pool. Both the actor and critic networks have a target network and a current network that have exactly the same structure. Each neural network is comprised of an input layer, output layer, two hidden layers, and a LSTM layers.

	The actor network  is responsible for generating the cache action  according to the observed state.
	\begin{equation}\label{eq11}
		\mathbf{a}_i= \mu(\mathbf{s}_i|\boldsymbol{\theta}^{a}) + \eta ^{i}.
	\end{equation} 
	where $\eta^{i}$ is  a random Gaussian variable that decreases with the training steps to allow the agent to explore potential better caching strategy. 
	
	The primary function of the critic network is to assess the actions performed by the actor network. This is accomplished by learning a state-action value function  $\mathcal{Q}(\mathbf{s},\mathbf{a}|\boldsymbol{\theta}^{c}) = E\{\sum_{i= 0}^{\infty} \gamma \times r_{i}\}$ that can be computed recursively using the Bellman equation,
	\begin{equation}\label{eq17}
		\mathcal{Q}(\mathbf{s}_i,\mathbf{a}_i|\boldsymbol{\theta}^{c}) =E\{r_{i}(\mathbf{s}_{i},\mathbf{a}_{i})+\gamma E\{\mathcal{Q}(\mathbf{s}_{i+1},\mathbf{a}_{i+1}|\boldsymbol{\theta}^{c})\}\},
	\end{equation}
	where $\gamma$ represents the discount factor utilized to indicate the extent of future impact exerted by the current actions.
	
	The experience replay technique involves sampling a mini-batch of transitions from the episode experience pool. Using these sampled transitions, the actor applies the policy gradient to update the current network parameters $\boldsymbol{\theta}^{a}$
	\begin{equation}\label{eq28}
		\nabla_{\boldsymbol{\theta}^{a}}J(\boldsymbol{\theta}^{a}) \approx  E\lbrack\nabla_{\boldsymbol{\theta}^{a}}\mathcal{Q}(\mathbf{s}_{i},\mathbf{a}_{i}|\boldsymbol{\theta}^{c})\nabla_{\boldsymbol{\theta}^{a}}\mu(\mathbf{s}_i,\boldsymbol{\theta}^{a})\rbrack,
	\end{equation}
	and the critic adjusts parameter $\boldsymbol{\theta}^{c}$ by minimizing the loss function
	\begin{equation}\label{eq29}
		Loss(\boldsymbol{\theta}^{c})=E\lbrack(\mathcal{Y}-\mathcal{Q}(\mathbf{s}_i,\mathbf{a}_i|\boldsymbol{\theta}^{c}))^2],
	\end{equation}
	and
	\begin{equation}\label{eq30}
		\mathcal{Y} = r_{i} + \gamma\acute{\mathcal{Q}}(\mathbf{s}_{i+1},\acute{\mu}(\mathbf{s}_{i+1}|\acute{\boldsymbol{\theta}^{a}}),\acute{\boldsymbol{\theta}^{c})},
	\end{equation}
	where LSTM-DDPG utilizes the two network technology to improve the stabilization of the actor and critic, $\mathcal{Y}$ and  $\acute{\mu}(\mathbf{s}_{\tau+1}|\acute{\boldsymbol{\theta}^a})$ are  the  output of the target network of actor and critic, respectively.
	
	In addition, the parameters of the target networks $\acute{\boldsymbol{\theta}^{a}},\acute{\boldsymbol{\theta}^{c} }$ can be softly updated according to the real-time updated parameter $\boldsymbol{\theta}^{a}$ and  $\boldsymbol{\theta}^{c}$, respectively,
	\begin{equation}\label{eq31}
		\boldsymbol{\theta}^{a'} \leftarrow \zeta\boldsymbol{\theta}^{a}+(1-\zeta)\boldsymbol{\theta}^{a'},
	\end{equation}
	\begin{equation}\label{eq32}
		\boldsymbol{\theta}^{c'} \leftarrow \zeta\boldsymbol{\theta}^c+(1-\zeta)\boldsymbol{\theta}^{c'},
	\end{equation}
	where $\zeta$ is the soft update factor. The details of LSTM-DDPG are shown in Algorithm 1.
	\begin{algorithm}
		\caption{Training Process of the LSTM-DDPG}\label{alg_ddpg}
		\SetKwData{Left}{left}\SetKwData{This}{this}\SetKwData{Up}{up}
		\SetKwFunction{Union}{Union}\SetKwFunction{FindCompress}{FindCompress}
		\SetKwInOut{Input}{Input}\SetKwInOut{Output}{Output}
		
		\Input{The training episode numbers $S$, the training steps $I\times K$, the discount factor $\gamma$, the soft update factor $\zeta$, the replay buffer $D$, the mini-batch sampling size $N$, the critic net learning rate $l^c$, the actor net learning rate $l^a$,
			the Gaussian noise  $\eta^{\tau}$ }
		\Output{ The actor networks' weights $\boldsymbol{\theta}^{a}$}
		\BlankLine
		{Initialize :  The actor network $\mu$, the critic network $\mathcal{Q}$ with weights $\theta^a$ and $\theta^c$}\;
		\For{$j\leftarrow 1$ \KwTo $S$}{
			Reset the environment, get the initial state\;
			\For{$i\leftarrow 1$ \KwTo $I$}{
				The agent makes the caching action $\mathbf{a}$ by observing the state $\mathbf{s}$, executes the action\;
				\For{$k\leftarrow 1$ \KwTo $K$}{
					Obtain the Off-RA decisions using Algorithm 2\;
                       Update the state of environment by  using formulas  ( \ref{eq18}, \ref{eq19})\;
					\If{$k == K-1$}{
						Get a reward $r$, transform to the new state $\acute{\mathbf{s}}$ according to formula (\ref{eqlstm})\; Store the transition  $(\mathbf{s}^j_{i},\mathbf{a}^j_{i},r^j_{i},\mathbf{s}^j_{i+1})$ into  buffer $D$\;
					}
				}
			 
			} 
		}
		\If{$j> N$}{
			Sample a  mini-batch experiences from $D$\;
			Update the actor network by using the policy gradient (\ref{eq28})\;
			Update the critic network by minimizing the loss function (\ref{eq29})\;
			Softly update the parameters of the target network in both the actor and the critic networks by formulas (\ref{eq31}, \ref{eq32})\;
		}
	\end{algorithm}\DecMargin{1em}
	\subsection{ Improved-GA based Off-RA Strategy}\label{cache}
	If the service caching policy $\mathbf{c}_i^{\star}$ is given, the problem can be reduced to a series of one-shot collaborative offloading and computing and bandwidth resource allocation (Off-RA) problem 
	\begin{small}
		\begin{alignat}{2}\label{p2}
			\max_{\mathbf{d}(k),\mathbf{b}(k),\mathbf{f}(k)}\quad &\sum_{n = 1}^{N}Q_{n,i}(k)
			\\
			\mbox{s.t.} \quad
			& d_{n,i}^m(k) \le \lceil f_{n,i}^m(k) \rceil \le c_{m}^{f\star}(i),\forall n \in \mathcal{N},\forall m \in \mathcal{M} &\tag{34a}\\
			&b_{n,i}^m(k)\le (1-d_{n,i}^0(k)),  \forall n\in\mathcal{N},\forall m\in\mathcal{M}&\tag{34b}\\
			&\sum_{m\in \mathcal{M}\cup\{0,C\}}d_{n,i}^m(k) = 1,\forall n\in\mathcal{N} &\tag{34c}\\
			&\sum_{n\in\mathcal{N}}{d_{n,i}^m(k) f_{n,i}^m(k)}\le 1, \forall m\in\mathcal{M}&\tag{34d}\\
			&\sum_{n\in\mathcal{N}}{b_{n,i}^m(k)}\le 1, \forall m\in\mathcal{M}&\tag{34e}\\
			&E_{n,i}(k)\le E_{n}^{th}, \forall n\in\mathcal{N}&\tag{34f}\\
			&T_{n,i}(k)\le T_{n}^{th}, \forall n\in\mathcal{N}&\tag{34g}\\
			&d_{n,i}^m(k) \in \{0,1\},\forall n\in\mathcal{N}, \forall m\in\mathcal{M}\cup\{0,C\}&\tag{34h}  \\
			&f_{n,i}^m(k)\in[0,1], \forall n\in\mathcal{N}, \forall m\in\mathcal{M}&\tag{34i}\\
			&b_{n,i}^m(k) \in[0,1],\forall n\in\mathcal{N},\forall m\in\mathcal{M}&\tag{34j}
		\end{alignat}
	\end{small}However, this is an  instantaneous one-step MINLP, which cannot be converted into MDP and thus cannot be solved by RL algorithms. To solve it, we introduce the GA, which is widely used to solve single-step resource optimization problems as formulated in our small timescale slots with remarkable performance. And it can be easily extended and embedded into DRL algorithms \cite{Zhang2022}. To adopt the GA, we  design a hybrid coding and cross-mutation scheme for solving the problem with both integer and continuous variables and redesign the fitness function to ensure that the constraints are not violated and proposed the Improved GA. The details of the Improved-GA based Off-RA algorithm are shown in the following.

	\emph{1) Chromosome and hybird coding}.
	The chromosome expression of an individual represents a solution to the Off-RA problem, which consists of three parts. To satisfy the constraints (34h, 34i, 34j), the chromosome is encoded to a discrete integer part and two continuous real value parts. For example, we denote the chromosome of individual $j$ in the  $l$th generation as $g_j^l(k) =\{0,1,\cdot,1;0,0.3,\cdot,0.2;0,0.2,\cdot,0.3\}_{N\times 3}$. 

	\emph{2) Fitness function}.
	The fitness function represents the quality of an individual's resource allocation strategy, with higher fitness indicating better performance. To construct the fitness function, we use the objective function of the total QoS of TDs. However, during the process of crossover and mutation, it is possible to generate individuals that violate constraints. To ensure the feasibility of the solution, we design the fitness function in a piecewise form. Specifically, if an individual violates the constraints, its fitness is defined as the minimum possible value. The fitness function is defined as follows
	\begin{equation}\label{34}
		F_{g_j^l}(k)=
		\begin{cases}
			e^{-3} & \text{$violation$}\\
			\sum_{n = 1}^{N}Q_{n,i}(k)  & \text{$else$}
		\end{cases}.
	\end{equation}
	\emph{3) Selection, Crossover, and Mutation }
	To preserve the excellent chromosomes in the current population, we introduce the fitness proportionate selection operation. Following the principle of survival of the fittest, certain chromosomes are selected as parent nodes for the next generation. The probability that the $j$th chromosome $g_j^l$ in the $l$th generation is selected as a parent node defined as
	\begin{equation}\label{35}
		\mathbf{PR}_{g_j^l}(k) =\frac{F_{g_j^l}(k)}{\sum_{g^l\in \mathcal{G}^l} F_{g^l}(k)}.
	\end{equation}
	After selection, parent nodes generate offspring through crossover and mutation. However, the original uniform crossover operation is unsuitable for consecutively coded chromosomes. As an alternative to direct swapping, we take the mean of the parents. Furthermore, for our problem, this type of crossover operation may not fully capture the coupling between variables. To address this, we add a self-crossover operation after the original uniform crossover. This operation randomly swaps genes in each block of discrete and continuous genes separately.
	To obtain the final chromosome of the next generation, a mutation operation is required with a specific probability. Integer random mutation is applied to the discrete gene block, whereas real Gaussian mutation is applied to the continuous gene block. For instance, if the parents have the chromosome $\{0,1,2,0.2,0.4.0.3,0.1,0.2,0.6\}$, when their $(1,5,7)$th genes are altered, the chromosome of offspring is $\{2,1,1,0.2,0.45.0.3,0.05,0.2,0.6\}$. The detailed process is summarized in Algorithm \ref{alg_small}, and the framework is shown in Fig. \ref{two-timescale}.
	\begin{algorithm}
		\caption{Improve GA-based Off-RA Algorithm}\label{alg_small}
		\SetKwData{Left}{left}\SetKwData{This}{this}\SetKwData{Up}{up}
		\SetKwFunction{Union}{Union}\SetKwFunction{FindCompress}{FindCompress}
		\SetKwInOut{Input}{Input}\SetKwInOut{Output}{Output}
		
		\Input{$c^{\star}$,$L$, $P$, $G$, $p_c$, $p_m$}
		\Output{$g^{\ast}(k)$}
		\BlankLine
		{Randomly generate initial population $\mathcal{G}^{1}(k)$}\;
		\For{$l\leftarrow 1$ \KwTo $L$}{
			\For{$j\leftarrow 1$ \KwTo $|\mathcal{G}|$}{
				Encode each chromosome $g^l_j(k)$ hybridly \;
				Calculate the fitness of each individual $g^l_j(k)$ using  formula (\ref{34})\;
			}
			
			Select excellent parent nodes $\mathcal{P}^l(k)$ from population $\mathcal{G}^l(k)$ based on probability (\ref{35})\;
			\For{$x\leftarrow 1$ \KwTo $P^l$}{
				Select another parent node from $\mathcal{P}^l$ randomly and perform original uniform crossover option with probability $p_c$\;
				Perform self-crossover operation with probability $p_c$\;
				\For{$d\leftarrow 1$ \KwTo $G$}{
					Perform mutation operation in each gene with probability $p_m$\;
				}
			}
			$\mathcal{G}^{l+1} \leftarrow  \mathcal{G}^{l}$
		}
		$g^{\ast} \leftarrow Max (\mathcal{G}^{L})$
	\end{algorithm}\DecMargin{1em}
	
	\subsection{ Computational Complexity of the proposed algorithms}
	For the proposed LSTM-DDPG based service caching algorithm, the training process runs offline in the CS where there are sufficient computing resources. Therefore, we focus only on the complexity of the online service caching process. For the LSTM-DDPG agent, only the actor network works online. We take the number of parameters in the model to represent the complexity of the algorithm. The actor network has an input layer, an output layer, two hidden layers, and a LSTM layer. The complexity of the actor network can be written as $O(N_i^aL_1+ 4(L_1L_2 +L_2^2+L_2)+L_2N_o^a)$, where $N_i^a$ is the number of neurons in the input layer, $N_o^a$ is the number of neurons in the output layer, $L_1,L_2$ are the number of neurons in the hidden layers.
	
	The complexity of  Algorithm \ref{alg_small} lies in the design of coding, crossover and  mutation options, but it is more directly limited by the generations and size of the population. Suppose the population size is P and the size of the chromosome is $G= 3\times N $ and the number of generations is L, the computational complexity is about  $O(2L+7LP+ 3LPN)$.
	
	\section{Simulation Results and Analysis}
	To validate the efficiency of the proposed DGL-DDPG scheme, we first verify the convergence
of the DGL-DDPG algorithm, and then we verify the long-term and short-term system performance
of the proposed algorithm under different environment settings. To investigate the performance in detail, we define the objective function, namely the weighted sum of all TD’s QoS and cache switching costs of all ESs, as the long-term system performance evaluation
metric $U_{large}$, and  define the average QoS of all TDs in a small timescale slot as the  short-term performance
metric. In detail, they are expressed as
\begin{equation}\label{40}
	U_{large} =\sum_{i=1}^{I}(\sum_{k=1}^{K}\sum_{n = 1}^{N} 
		Q_{n,i}(k) - \delta Cost(i)).
	\end{equation}
\begin{equation}\label{40}
		U_{small} = \frac{1}{N}\sum_{n=1}^{N} Q_{n,i}(k).
	\end{equation}

	\begin{table}[t]
		\caption{Simulation parameters}\label{tab1}
		\small
		\centering
		\begin{tabular}{|c|c|}
			\hline
			Parameters   & Values   \\
			\hline
			Number of ESs $M$ & [2,5]\\
			\hline
			Number of TDs $N$ & [20,50]\\
			\hline
			Input size of task $s^{input}$  & [500,5000]KB \\
			\hline
			Cache size of task $s^{cache}$& [2,20]GB\\
			\hline
			Computing frequency of task & [400,1000]$cycle/bit$\\
			\hline
			Energy coefficient of TD  $\varepsilon$  & $5\times10^{-27}$\\
			\hline
			Computing capacity of CS $F^c$ & 5GHz\\
			\hline
			Computing capacity of TD $F^l$ & 1GHz\\
			\hline
			Weight factor $\alpha,\beta$ & 0.5,0.5\\
			\hline
			Balance factor $\delta$ & 0.001\\
			\hline
			Size of each generation $P$  & 30\\
			\hline
			Crossover rate $p_c$& 0.45\\
			\hline
			Mutation rate $p_m$ & 0.1\\
			\hline
			The soft update factor $\zeta$  &0.05\\
			\hline
			Actor network learning rate $\l_a$ &0.001\\
			\hline
			Critic network learning rate $\l_c$ &0.002\\
			\hline
			Discount factor $\gamma$ &0.95\\
			\hline
			Memory pool capacity $D$ &10000\\
			\hline
			Mini-batch size  &32\\
			\hline
		\end{tabular}
	\end{table}
	
	\subsection{System Setups}
	We consider a scenario consisting of one CS, $M$ ESs, and $N$ TDs within a 500m $\times$ 500m square area. The system has 10 types of tasks and services, each TD generates one task randomly per small-timescale slot according to a Zipf distribution with a parameter of 0.8.
	For each ES, the available bandwidth, computing, and caching resource are set to 20MHz, 20GHz, and 50GB, respectively. The TD's transmission power and computing capacity are set to 20dBm and 1GHz, respectively. The fixed transmission rates  $R^{back}$ and $R^{co}$  are set to be 200Mbps and 20Mbps, respectively. The additive white Gaussian noise $\sigma^2$ is set to 114dBm, with a path loss exponent of 2. Table 1 summarizes the rest simulation parameters.
	To show the priority of  DGL-DDPG algorithm,  seven similar  algorithms are adopted as  baselines. 
	\begin{itemize}
		\item DDPG-based service caching (DDPG), where the service caching is optimized based on the traditional DDPG algorithm, and the Off-RA strategy is determined by using the proposed Improved-GA.
		\item Genetic algorithm (GA), where the service caching, offloading, bandwidth, and computing resource allocation decisions are generated by  traditional GA. This approach employs discrete coding scheme.
		\item No-cooperation, the  No-cooperation scheme does not consider the collaborative computing and caching among ESs. If the associated ES has related services, the TD can offload, otherwise, it will be processed locally or offloaded to the CS.
		\item Ave-resource, the communication and computing resources are  allocated equally to TDs, while the service caching and task offloading are optimized based on our scheme.
		\item Popular-cache, the Popular caching scheme caches service models at ESs based on statistical popularity information, while Off-RA decisions are made based on our proposed scheme.
		\item Random-cache, the Random caching scheme caches service models randomly at ESs under the constraint of cache capacity, while Off-RA decisions are made based on our scheme.
		\item Random-off, the Random offloading scheme generates offloading decisions randomly, while the other decisions are made based on our proposed scheme.
	\end{itemize}
	
	\subsection{The Convergence Property of  DGL-DDPG Algorithm}
 The proposed algorithm consists of both the large-timescale LSTM-DDPG based service caching and the small-timescale Improved-GA based Off-RA.
	The service caching strategy is learned by the LSTM-DDPG agent, and the agent gives a service caching decision at the beginning of each large-timescale slot, and then the Off-RA policy is obtained by the Improved-GA. First of all, we make sure that the  small-timescale Improved-GA based Off-RA  converges  for any given service caching strategy. It is then embedded into the large timescale LSTM-DDPG based service caching to train the LSTM-DDPG agent.  Therefore, when  the large-timescale LSTM-DDPG converges, it means that the designed two-timescale DGL-DDPG algorithm converges. So, in the following, we first analyze the convergence of the small-timescale Improved-GA based Off-RA, and then analyze the convergence of the large-timescale LSTM-DDPG based service caching.

\begin{figure}
		\centering
		\includegraphics[width=0.45\textwidth]{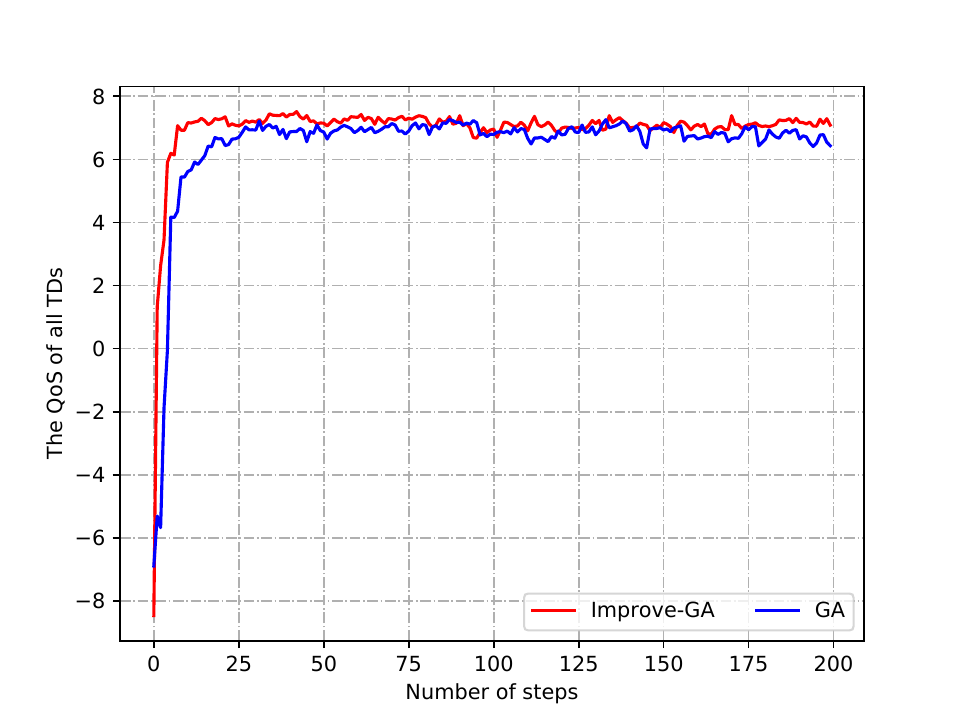}\\
		\caption{Convergence performance of Improved-GA.}\label{GA}
	\end{figure}

The convergence property of the Improved-GA and GA  is  shown in Fig. \ref{GA}. It is seen that Improved-GA converges faster, and  converges within about 10 steps, while  GA converges within about 25 steps. This improvement is brought by the implementation of hybrid coding and self-crossover operations, which require less exploration steps but yield superior results. 
This also reduces the training complexity and  time  of the large-timescale LSTM-DDPG.
Therefore, in the training process of LSTM-DDPG agent, we set the iteration number to be 20 steps.

\begin{figure}
		\centering
		\includegraphics[width=0.45\textwidth]{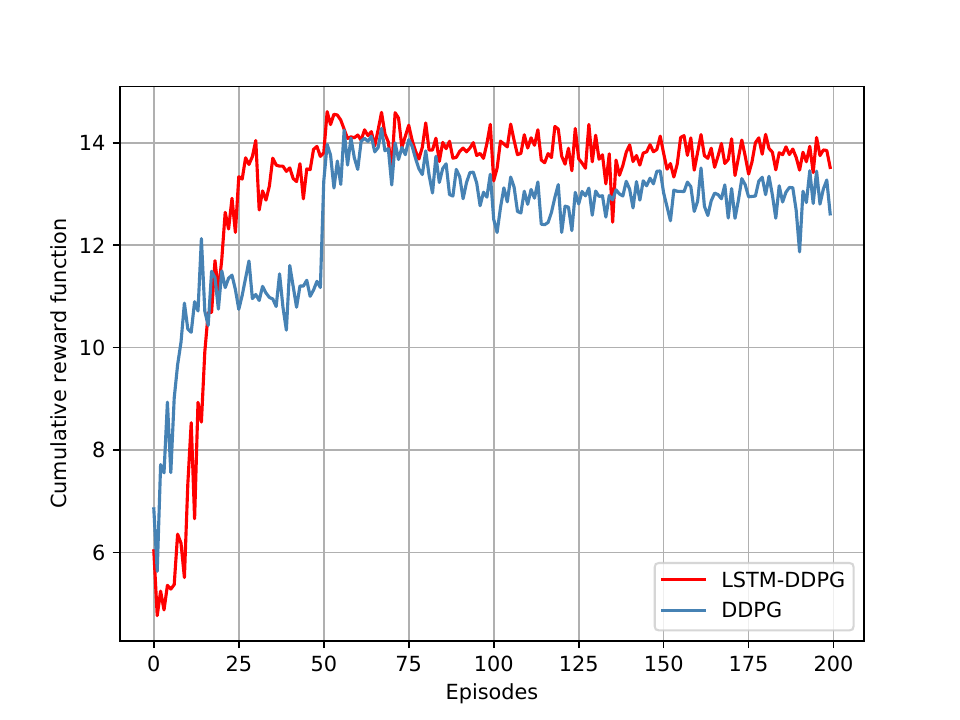}\\
		\caption{Convergence performance of accumulative reward for LSTM-DDPG.}\label{lstmddpg}
	\end{figure}

	Fig. \ref{lstmddpg} shows the convergence property of  LSTM-DDPG  and DDPG. It is observed that LSTM-DDPG performs  better and more stable  than DDPG. In the early stages of experience accumulation and exploration, both the two curves grow rapidly, indicating that DRL algorithms can efficiently acquire policies by interacting with their corresponding environments.
	As the number of training episodes increases, both the two algorithms converge, the difference is that LSTM-DDPG is faster and a bit more stable. This is because that the LSTM-based network structure can better leverage historical information to predict unknown network observation transitions.
	Effective predictions reduce the number of invalid explorations and therefore make it converge faster.

\subsection{The Performance of  DGL-DDPG at  large-timescale}
	To fully verify the efficiency of  DGL-DDPG in dealing with the two-timescale problem, we compare its long-term system performance, $
		U_{large}$, with that of DDPG and GA algorithm under different environment settings. The simulation is all in the setting of $I=20,K=5,M=2,N=20$ and  the result takes 100 averaging operations.

Fig. \ref{cache} presents the long-term utility of DGL-DDPG compared to other benchmarks under different caching capacity settings. When the caching capacity is set to a low value (20GB-60GB), the long-term utility of DGL-DDPG is significantly better than those of  GA and Popular-cache. 
This is because that GA only optimize single-step cache decisions based on the current offloading and resource allocation state,  while Popular-cache  always cache a few fixed, highly popular service models.  As a result, these benchmarks  fail to adapt to dynamic offloading environments and result in poor long-term performance.
In the same time, when achieving the same performance, DRL-based schemes occupy less caching resources than instantaneous GA and statistics-based Popular-cache.
Moreover, the performance gap between DGL-DDPG and DDPG further validates the effectiveness of adopting LSTM into DDPG for the long-term service caching phase.

\begin{figure}
		\centering
		\includegraphics[width=0.45\textwidth]{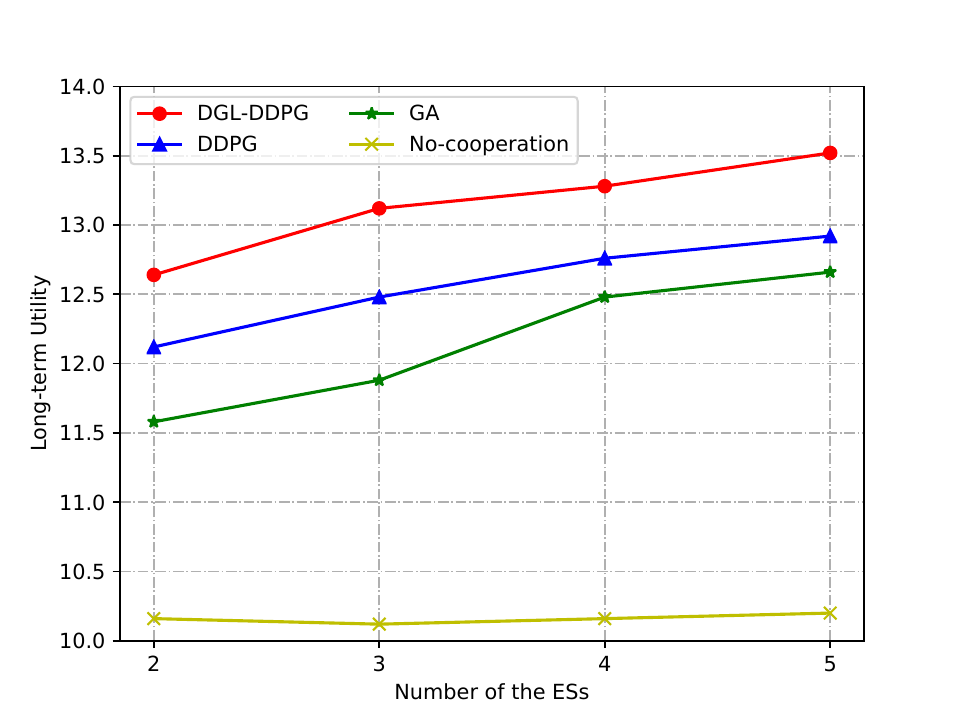}\\
		\caption{The long-term utility versus the number of  collaborative ES.}\label{cooperative}
	\end{figure}

The long-term utility for different number of the ES is shown in Fig. \ref{cooperative}, where No-cooperation  is  used as a benchmark. Apparently, the utility will increase as the number of the ESs increases.  The reason is that the whole system can schedule more resources.
Both DGL-DDPG and DDPG  outperform GA, demonstrating the superiority of DRL in  adapting to dynamic environment and solving the long-term optimization problem,
In addition, that DGL-DDPG  performs better than DDPG indicating that the proposed DGL-DDPG is more suited for solving the resource allocation problem  over two-timescale, since the LSTM network can enhance the temporal decision-making ability of DDPG.

	\begin{figure}
		\centering
		\includegraphics[width=0.45\textwidth]{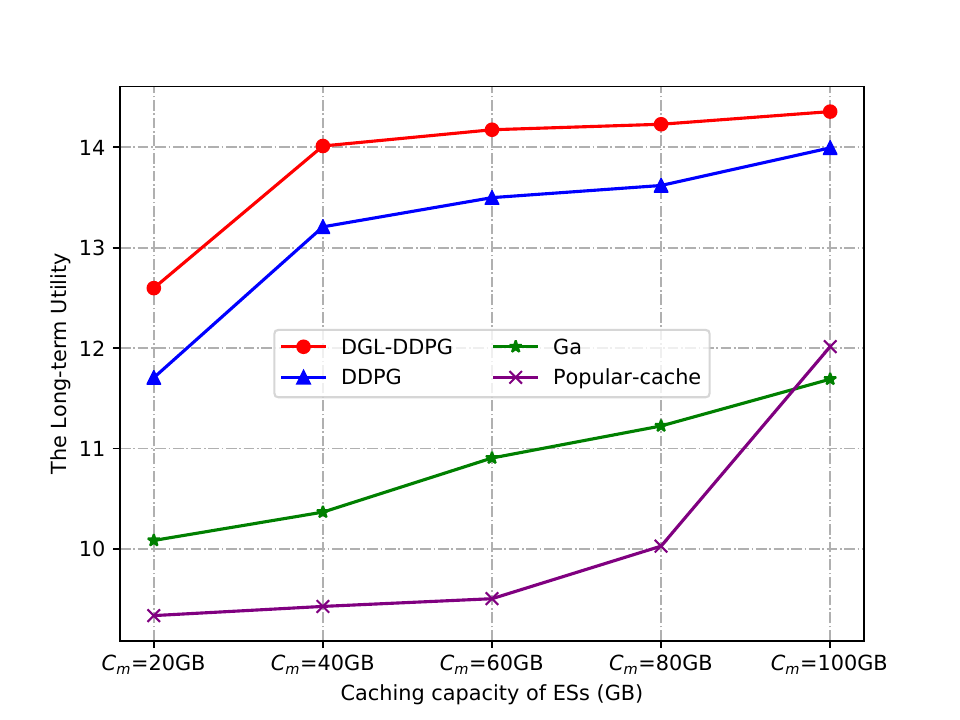}\\
		\caption{The long-term utility versus the cache capacity of ESs.}\label{cache}
	\end{figure}
	
\subsection{The Performance of DGL-DDPG  at Small-timescale}
In this part, we compare the instantaneous performance $U_{small}$ of the DGL-DDPG algorithm at small-timescale slot.

\begin{figure}
		\centering
		\includegraphics[width=0.45\textwidth]{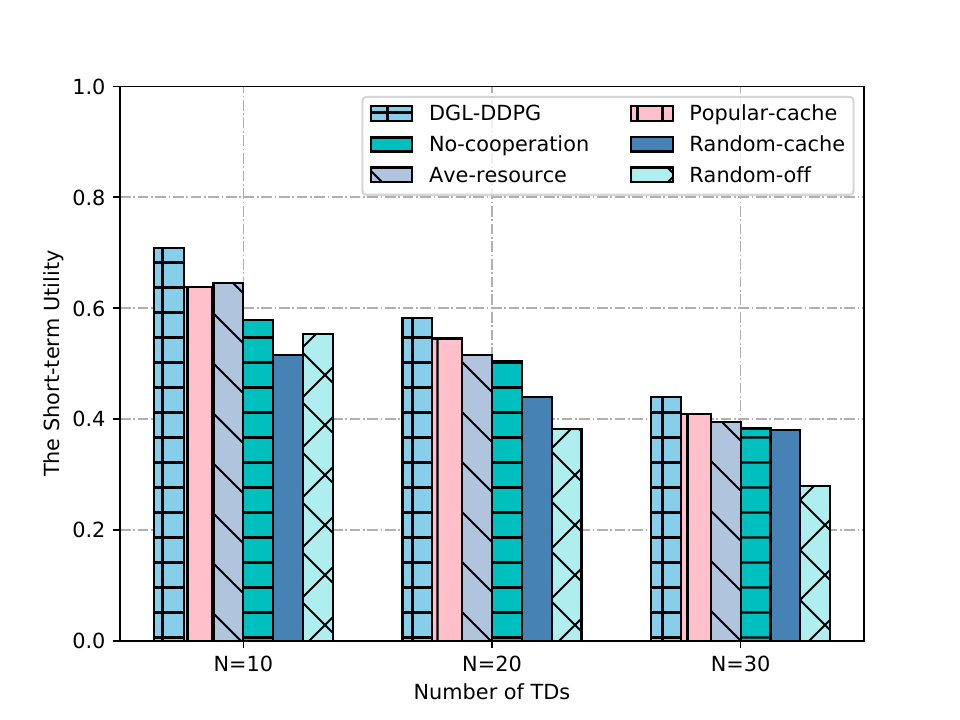}\\
		\caption{The short-term utility versus the number of TDs.}\label{schemeuser}
	\end{figure}

Fig. \ref{schemeuser} shows the impacts of the number of TDs on the short-term utility. As the number of TDs increases, the short-term utility of all schemes continuously decreases. This is because the total resources of the ESs are limited. However, DGL-DDPG still performs better than  Popular-cache and Ave-resource  when the number of TD is large. The performance advantage over No-cooperation, Random-cache and Random-off is more obvious. The reason is that  DGL-DDPG can achieve higher resource utilization through the form of collaboration and joint resource optimization.
	
\begin{figure}
		\centering
		\includegraphics[width=0.45\textwidth]{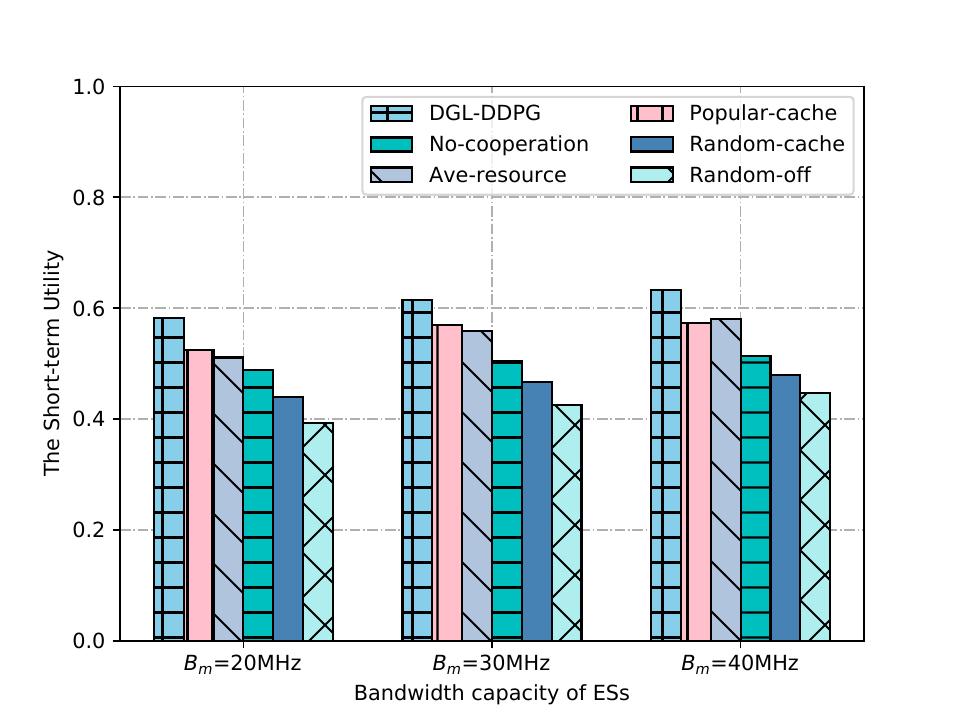}\\
		\caption{The  short-term utility versus the bandwidth resource of ESs.}\label{schemeband}
	\end{figure}

The short-term utility with different bandwidth resource at the ESs is shown in Fig. \ref{schemeband}. 
It is seen that the bandwidth resources increase, the short-term utility of all the simulated schemes improves,  and DGL-DDPG still performs the best.
This is because that  the bandwidth resources limit the number of the offloaded tasks, and an increment in bandwidth resources enables the ESs to handle more tasks. 
Moreover,  DGL-DDPG achieves a higher gain than  No-cooperation and Popular-cache schemes. It reveals that even if tasks can be offloaded to the ESs, they cannot be handled if there are no the corresponding service model and computing resources, which means it is necessary to schedule the multi-dimensional resources together. 
	
\begin{figure}
		\centering
		\includegraphics[width=0.45\textwidth]{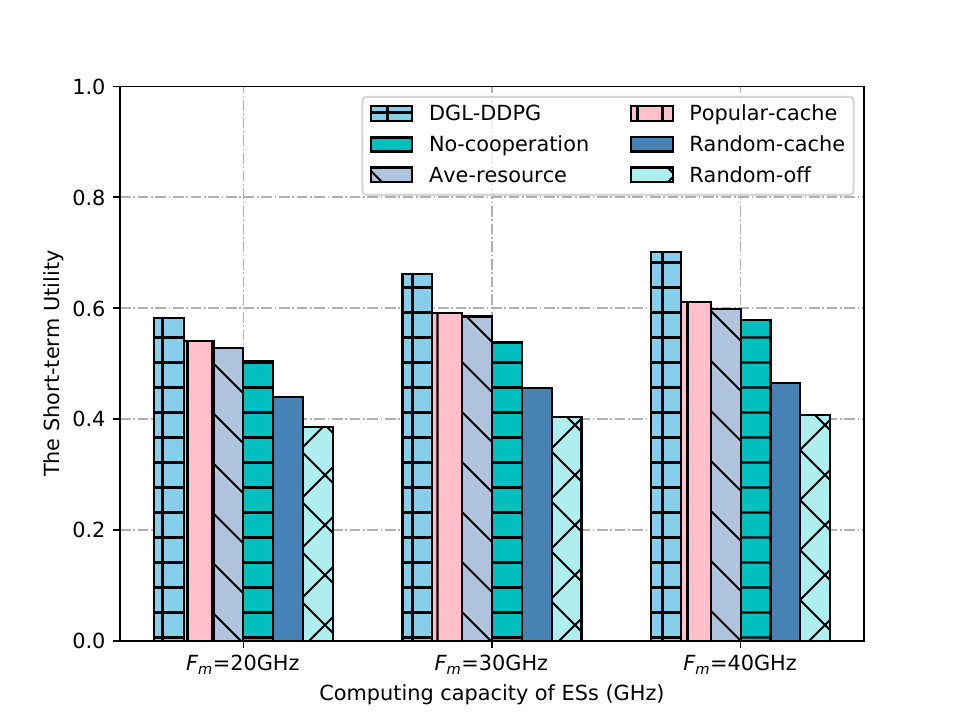}\\
		\caption{The  short-term utility versus the computing resource of ESs.}\label{schemecomp}
	\end{figure}

Fig. \ref{schemecomp} shows the impacts of the computing capacity of the ESs on the short-term utility.  It is seen that the short-term utility of all schemes improves as the computing capacity of the ESs increasing from 20GHz to 30 GHz.
The reason is  that more tasks can be processed at the ESs with the lower task processing delay  and energy consumption when there is more computing resources.
This reveals  that  DGL-DDPG realizes a reasonable service caching strategy along with the offloading decisions to make full use of the computing resource. Moreover, the gain compared to No-cooperation verifies the advantage of collaboration.
	
\section{Conclusions}
This article proposed a joint solution for service caching, collaborative offloading, computing, and bandwidth resources allocation for MEC-enabled IIoT systems with ESs cooperation. The system utility was defined by tasking account into the QoS of TDs and  the cache switching costs, based on which a long-term utility maximizing problem was formulated  under the constraints of limited resources.
The optimization  problem is two-timescale MINLP. To efficiently solve it, we proposed a two-timescale multi-dimensional resources allocation scheme, DGL-DDPG.
A LSTM-DDPG agent was designed to solve the service caching for the large timescale by mining the temporal relationship between long-term service caching and short-term Off-RA decisions. Then, an Improved-GA algorithm was embedded into the agent to make instantaneous Off-RA decisions by incorporating a hybrid coding and cross-mutation schemes.
	Extensive simulations were conducted, and the results had demonstrated the superiority of the proposed DGL-DDPG scheme in terms of the convergence, the average QoS of TDs  and  the long-term utility  through comparison with the other benchmarks under different system settings.

	\bibliographystyle{IEEEtranN}
	\bibliography{jont}

\begin{thebibliography}{41}
\providecommand{\natexlab}[1]{#1}
\providecommand{\url}[1]{#1}
\csname url@samestyle\endcsname
\providecommand{\newblock}{\relax}
\providecommand{\bibinfo}[2]{#2}
\providecommand{\BIBentrySTDinterwordspacing}{\spaceskip=0pt\relax}
\providecommand{\BIBentryALTinterwordstretchfactor}{4}
\providecommand{\BIBentryALTinterwordspacing}{\spaceskip=\fontdimen2\font plus
\BIBentryALTinterwordstretchfactor\fontdimen3\font minus \fontdimen4\font\relax}
\providecommand{\BIBforeignlanguage}[2]{{%
\expandafter\ifx\csname l@#1\endcsname\relax
\typeout{** WARNING: IEEEtranN.bst: No hyphenation pattern has been}%
\typeout{** loaded for the language `#1'. Using the pattern for}%
\typeout{** the default language instead.}%
\else
\language=\csname l@#1\endcsname
\fi
#2}}
\providecommand{\BIBdecl}{\relax}
\BIBdecl

\bibitem[Dai et~al.(2023)Dai, Xiao, Jiang, Alazab, Lui, Dustdar, and Liu]{dai2023task}
X.~Dai, Z.~Xiao, H.~Jiang, M.~Alazab, J.~C. Lui, S.~Dustdar, and J.~Liu, ``{Task co-offloading for D2D-assisted mobile edge computing in industrial internet of things},'' \emph{IEEE Trans. Industr. Inform.}, vol.~19, no.~1, pp. 480--490, Jan. 2023.

\bibitem[Chen et~al.(2022{\natexlab{a}})Chen, Sun, Yang, and Taleb]{Chen2022}
Y.~Chen, Y.~Sun, B.~Yang, and T.~Taleb, ``{Deep reinforcement learning-based joint caching and computing edge service placement for sensing-data-driven IIoT applications},'' in \emph{Proc. IEEE Int. Conf. Commun. (ICC)}, Seoul, South Korea, May 2022, pp. 4287--4292.

\bibitem[Qiu et~al.(2020)Qiu, Chi, Zhou, Ning, Atiquzzaman, and Wu]{Qiu2020}
T.~Qiu, J.~Chi, X.~Zhou, Z.~Ning, M.~Atiquzzaman, and D.~O. Wu, ``Edge computing in industrial internet of things: Architecture, advances and challenges,'' \emph{IEEE Commun. Surv. Tutor.}, vol.~22, no.~4, pp. 2462--2488, Fourthquarter 2020.

\bibitem[Zhang and Wei(2021)]{Zhang2021b}
Y.~Zhang and H.-Y. Wei, ``{Risk-aware cloud-edge computing framework for delay-sensitive industrial IoTs},'' \emph{IEEE Trans. Netw. Service Manag.}, vol.~18, no.~3, pp. 2659--2671, Sep. 2021.

\bibitem[Deng et~al.(2022{\natexlab{a}})Deng, Chen, Zhu, Fang, Chen, and Deng]{Deng2022}
Y.~Deng, X.~Chen, G.~Zhu, Y.~Fang, Z.~Chen, and X.~Deng, ``Actions at the edge: Jointly optimizing the resources in multi-access edge computing,'' \emph{IEEE Wirel. Commun.}, vol.~29, no.~2, pp. 192--198, Apr. 2022.

\bibitem[Liu et~al.(2022)Liu, Zhang, Ding, and Yuan]{Liu2022a}
W.~Liu, H.~Zhang, H.~Ding, and D.~Yuan, ``Delay and energy minimization for adaptive video streaming: A joint edge caching, computing and power allocation approach,'' \emph{IEEE Trans. Veh. Technol.}, vol.~71, no.~9, pp. 9602--9612, Sep. 2022.

\bibitem[Zhang et~al.(2021{\natexlab{a}})Zhang, Di, Zheng, Lin, and Song]{Zhang2021d}
Y.~Zhang, B.~Di, Z.~Zheng, J.~Lin, and L.~Song, ``Distributed multi-cloud multi-access edge computing by multi-agent reinforcement learning,'' \emph{IEEE Trans. Wirel. Commun.}, vol.~20, no.~4, pp. 2565--2578, April 2021.

\bibitem[Wang et~al.(2022)Wang, Di, Song, and Jennings]{Wang2022}
P.~Wang, B.~Di, L.~Song, and N.~R. Jennings, ``Multi-layer computation offloading in distributed heterogeneous mobile edge computing networks,'' \emph{IEEE Trans. Cogn. Commun. Netw.}, vol.~8, no.~2, pp. 1301--1315, Jun. 2022.

\bibitem[Wu et~al.(2022)Wu, Lin, and Quek]{Wu2022}
Y.-C. Wu, C.~Lin, and T.~Q.~S. Quek, ``{A robust distributed hierarchical online learning approach for dynamic MEC networks},'' \emph{IEEE J. Sel. Areas Commun.}, vol.~40, no.~2, pp. 641--656, Feb. 2022.

\bibitem[Yan et~al.(2021)Yan, Bi, Duan, and Zhang]{Yan2021}
J.~Yan, S.~Bi, L.~Duan, and Y.-J.~A. Zhang, ``Pricing-driven service caching and task offloading in mobile edge computing,'' \emph{IEEE Trans. Wirel. Commun.}, vol.~20, no.~7, pp. 4495--4512, Jul. 2021.

\bibitem[Chen et~al.(2022{\natexlab{b}})Chen, Gong, Jiang, Zhou, and Chen]{Chen2022a}
L.~Chen, G.~Gong, K.~Jiang, H.~Zhou, and R.~Chen, ``{DDPG-based computation offloading and service caching in mobile edge computing},'' in \emph{Proc. IEEE Conf. Comput. Commun. Workshops (INFOCOM WKSHPS)}, Virtual, May 2022, pp. 1--6.

\bibitem[Pham et~al.(2021)Pham, Nguyen, Nguyen, and Huh]{Pham2021}
X.-Q. Pham, T.-D. Nguyen, V.~Nguyen, and E.-N. Huh, ``{Joint service caching and task offloading in multi-access edge computing: A QoE-based utility optimization approach},'' \emph{IEEE Commun. Lett.}, vol.~25, no.~3, pp. 965--969, Mar. 2021.

\bibitem[Ko et~al.(2022)Ko, Kim, Jung, and Choi]{Ko2022}
S.-W. Ko, S.~J. Kim, H.~Jung, and S.~W. Choi, ``Computation offloading and service caching for mobile edge computing under personalized service preference,'' \emph{IEEE Trans. Wirel. Commun.}, vol.~21, no.~8, pp. 6568--6583, Aug. 2022.

\bibitem[Zhang et~al.(2021{\natexlab{b}})Zhang, Kishk, and Alouini]{Zhang2021f}
Y.~Zhang, M.~Kishk, and S.~Alouini, ``{Computation offloading and service caching in heterogeneous MEC wireless networks},'' \emph{IEEE Trans. Mob. Comput.}, pp. 1--1, Dec. 2021.

\bibitem[Zhang et~al.(2021{\natexlab{c}})Zhang, Zhang, Zhang, Shen, and Wang]{Zhang2021a}
G.~Zhang, S.~Zhang, W.~Zhang, Z.~Shen, and L.~Wang, ``Joint service caching, computation offloading and resource allocation in mobile edge computing systems,'' \emph{IEEE Trans. Wirel. Commun.}, vol.~20, no.~8, pp. 5288--5300, Aug. 2021.

\bibitem[Alqerm and Pan(2021)]{Alqerm2021}
I.~Alqerm and J.~Pan, ``{DeepEdge: A new QoE-based resource allocation framework using deep reinforcement learning for future heterogeneous edge-IoT applications},'' \emph{IEEE Trans. Netw. Service Manag.}, vol.~18, no.~4, pp. 3942--3954, Dec. 2021.

\bibitem[Kamran et~al.(2022)Kamran, Yeh, and Ma]{Kamran2022}
K.~Kamran, E.~Yeh, and Q.~Ma, ``{DECO: Joint computation scheduling, caching, and communication in data-intensive computing Nntworks},'' \emph{IEEE ACM Trans. Netw.}, vol.~30, no.~3, pp. 1058--1072, Jun. 2022.

\bibitem[Kai et~al.(2021{\natexlab{a}})Kai, Zhou, Yi, and Huang]{Kai2021a}
C.~Kai, H.~Zhou, Y.~Yi, and W.~Huang, ``Collaborative cloud-edge-end task offloading in mobile-edge computing networks with limited communication capability,'' \emph{IEEE Trans. Cogn. Commun. Netw.}, vol.~7, no.~2, pp. 624--634, June 2021.

\bibitem[Zhang et~al.(2021{\natexlab{d}})Zhang, Di, Zheng, Lin, and Song]{Zhang2021}
Y.~Zhang, B.~Di, Z.~Zheng, J.~Lin, and L.~Song, ``Distributed multi-cloud multi-access edge computing by multi-agent reinforcement learning,'' \emph{IEEE Trans. Wirel. Commun.}, vol.~20, no.~4, pp. 2565--2578, Apr. 2021.

\bibitem[Sun et~al.(2021)Sun, Wu, Li, Fan, Wen, and Leung]{Sun2021}
C.~Sun, X.~Wu, X.~Li, Q.~Fan, J.~Wen, and V.~C. Leung, ``{Cooperative computation offloading for multi-access edge computing in 6G mobile networks via soft actor critic},'' \emph{IEEE Trans. Netw. Sci. Eng.}, pp. 1--1, Apr. 2021.

\bibitem[Sahni et~al.(2020)Sahni, Cao, Yang, and Ji]{sahni2020multi}
Y.~Sahni, J.~Cao, L.~Yang, and Y.~Ji, ``Multi-hop multi-task partial computation offloading in collaborative edge computing,'' \emph{IEEE Trans. Parallel Distrib. Syst.}, vol.~32, no.~5, pp. 1133--1145, Dec. 2020.

\bibitem[Tan et~al.(2023)Tan, Kuang, Gao, and Zhao]{Tan2023}
L.~Tan, Z.~Kuang, J.~Gao, and L.~Zhao, ``Energy-efficient collaborative multi-access edge computing via deep reinforcement learning,'' \emph{IEEE Transactions on Industrial Informatics}, vol.~19, no.~6, pp. 7689--7699, 2023.

\bibitem[Chen et~al.(2022{\natexlab{c}})Chen, Kuang, and Zhao]{Chen2022b}
Q.~Chen, Z.~Kuang, and L.~Zhao, ``Multiuser computation offloading and resource allocation for cloud–edge heterogeneous network,'' \emph{IEEE Internet of Things Journal}, vol.~9, no.~5, pp. 3799--3811, March 2022.

\bibitem[Li et~al.(2022{\natexlab{a}})Li, Deng, Chen, Deng, and Yin]{li2022mec}
B.~Li, X.~Deng, X.~Chen, Y.~Deng, and J.~Yin, ``{MEC-based dynamic controller placement in SD-IoV: A deep reinforcement learning approach},'' \emph{IEEE Trans. Veh. Technol.}, vol.~71, no.~9, pp. 10\,044--10\,058, Jun. 2022.

\bibitem[Zhang et~al.(2022)Zhang, Yang, Shang, and Zhang]{Zhang2022}
H.~Zhang, Y.~Yang, B.~Shang, and P.~Zhang, ``Joint resource allocation and multi-part collaborative task offloading in mec systems,'' \emph{IEEE Transactions on Vehicular Technology}, vol.~71, no.~8, pp. 8877--8890, 2022.

\bibitem[Zhang et~al.(2023{\natexlab{a}})Zhang, Chen, Wang, and Zhu]{Zhang2023a}
J.~Zhang, S.~Chen, X.~Wang, and Y.~Zhu, ``Dynamic reservation of edge servers via deep reinforcement learning for connected vehicles,'' \emph{IEEE Trans. Mob. Comput.}, vol.~22, no.~5, pp. 2661--2674, May 2023.

\bibitem[Li et~al.(2022{\natexlab{b}})Li, Zhang, Yuan, and Zhang]{Li2022}
D.~Li, H.~Zhang, D.~Yuan, and M.~Zhang, ``Learning-based hierarchical edge caching for cloud-aided heterogeneous networks,'' \emph{IEEE Trans. Wirel. Commun.}, pp. 1--1, Sep. 2022.

\bibitem[Li et~al.(2023)Li, Zhang, Ding, Li, Liang, and Yuan]{Li2023}
D.~Li, H.~Zhang, H.~Ding, T.~Li, D.~Liang, and D.~Yuan, ``{User preference learning-based proactive edge caching for D2D-assisted wireless networks},'' \emph{IEEE Internet Things J.}, pp. 1--1, Feb. 2023.

\bibitem[Zhang et~al.(2021{\natexlab{e}})Zhang, Chen, Wang, and Zhu]{Zhang2021g}
J.~Zhang, S.~Chen, X.~Wang, and Y.~Zhu, ``Deepreserve: Dynamic edge server reservation for connected vehicles with deep reinforcement learning,'' in \emph{Proc. IEEE Conf. Comput. Commun. (INFOCOM)}, Virtual, May 2021, pp. 1--10.

\bibitem[Liu and Cao(2021)]{Liu2021}
H.~Liu and G.~Cao, ``Deep reinforcement learning-based server selection for mobile edge computing,'' \emph{IEEE Trans. Veh. Technol.}, vol.~70, no.~12, pp. 13\,351--13\,363, Dec. 2021.

\bibitem[Tan et~al.(2022)Tan, Kuang, Zhao, and Liu]{Tan2022}
L.~Tan, Z.~Kuang, L.~Zhao, and A.~Liu, ``{Energy-efficient joint task offloading and resource allocation in OFDMA-based collaborative edge computing},'' \emph{IEEE Trans. Wirel. Commun.}, vol.~21, no.~3, pp. 1960--1972, Mar. 2022.

\bibitem[Kai et~al.(2021{\natexlab{b}})Kai, Zhou, Yi, and Huang]{Kai2021}
C.~Kai, H.~Zhou, Y.~Yi, and W.~Huang, ``Collaborative cloud-edge-end task offloading in mobile-edge computing networks with limited communication capability,'' \emph{IEEE Trans. Cogn. Commun. Netw.}, vol.~7, no.~2, pp. 624--634, Jun. 2021.

\bibitem[Deng et~al.(2022{\natexlab{b}})Deng, Chen, Chen, and Fang]{Deng2022a}
Y.~Deng, Z.~Chen, X.~Chen, and Y.~Fang, ``Throughput maximization for multiedge multiuser edge computing systems,'' \emph{IEEE Internet Things J.}, vol.~9, no.~1, pp. 68--79, Jan. 2022.

\bibitem[Chu et~al.(2022)Chu, Yu, Yu, Lui, and Lin]{Chu2022}
W.~Chu, P.~Yu, Z.~Yu, J.~C. Lui, and Y.~Lin, ``Online optimal service selection, resource allocation and task offloading for multi-access edge computing: A utility-based approach,'' \emph{IEEE Trans. Mob. Comput.}, pp. 1--1, Feb. 2022.

\bibitem[Zhou et~al.(2022{\natexlab{a}})Zhou, Zhang, Wu, Dong, and Leung]{Zhou2022a}
H.~Zhou, Z.~Zhang, Y.~Wu, M.~Dong, and V.~C.~M. Leung, ``Energy efficient joint computation offloading and service caching for mobile edge computing: A deep reinforcement learning approach,'' \emph{IEEE Trans. Green Commun. Netw.}, pp. 1--1, Jul. 2022.

\bibitem[Zhou et~al.(2023)Zhou, Wang, Zheng, He, and Dong]{Zhou2023}
H.~Zhou, Z.~Wang, H.~Zheng, S.~He, and M.~Dong, ``{Cost minimization-oriented computation offloading and service caching in mobile cloud-edge computing: An A3C-based approach},'' \emph{IEEE Trans. Netw. Sci. Eng.}, pp. 1--12, Mar. 2023.

\bibitem[Zhang et~al.(2023{\natexlab{b}})Zhang, Shen, Wang, Zhang, and Wang]{Zhang2023}
J.~Zhang, Y.~Shen, Y.~Wang, X.~Zhang, and J.~Wang, ``{Dual-timescale resource allocation for collaborative service caching and computation offloading in IoT systems},'' \emph{IEEE Trans. Industr. Inform.}, vol.~19, no.~2, pp. 1735--1746, Feb. 2023.

\bibitem[Zhou et~al.(2022{\natexlab{b}})Zhou, Wu, Tan, and Zhang]{Zhou2022b}
R.~Zhou, X.~Wu, H.~Tan, and R.~Zhang, ``{Two time-scale joint service caching and task offloading for UAV-assisted mobile edge computing},'' in \emph{Proc. IEEE Conf. Comput. Commun. (INFOCOM)}, Virtual, May 2022, pp. 1189--1198.

\bibitem[Bi et~al.(2020)Bi, Huang, and Zhang]{Bi2020}
S.~Bi, L.~Huang, and Y.-J.~A. Zhang, ``Joint optimization of service caching placement and computation offloading in mobile edge computing systems,'' \emph{IEEE Trans. Wirel. Commun.}, vol.~19, no.~7, pp. 4947--4963, Jul. 2020.

\bibitem[Wen et~al.(2020)Wen, Cui, Quek, Zheng, and Jin]{Wen2020}
W.~Wen, Y.~Cui, T.~Q.~S. Quek, F.-C. Zheng, and S.~Jin, ``Joint optimal software caching, computation offloading and communications resource allocation for mobile edge computing,'' \emph{IEEE Trans. Veh. Technol.}, vol.~69, no.~7, pp. 7879--7894, Jul. 2020.

\bibitem[Sun et~al.(2022)Sun, Chen, Wang, and Mao]{Sun2022}
Y.~Sun, S.~Chen, Z.~Wang, and S.~Mao, ``A joint learning and game-theoretic approach to multi-dimensional resource management in fog radio access networks,'' \emph{IEEE Trans. Veh. Technol.}, pp. 1--14, Oct. 2022.

\end{thebibliography}
\end{document}